\documentclass[
 reprint,
superscriptaddress,
 amsmath,amssymb
 aps, nofootinbib
]{revtex4-2}

\usepackage{diagbox}
\usepackage{multirow}
\usepackage{float}
\usepackage[linesnumbered, ruled,vlined]{algorithm2e}
\usepackage{tikz}
\usepackage{quantikz}
\usepackage{color}
\usepackage{amsmath}
\usepackage{amssymb}
\usepackage{graphicx}
\usepackage[caption=false]{subfig}
\usepackage{bm}% bold math
\usepackage[hidelinks]{hyperref}% add hypertext capabilities
\usepackage{textpos}
\usepackage{wrapfig}
\usepackage{footnote}
\usepackage{cleveref}
\usepackage{cancel}
\usepackage{paralist}

\newtheorem{definition}{Definition}[section]

\newtheorem{lemma}{Lemma}[section]

\renewcommand{\bra}[1]{\langle{#1}|}
\renewcommand{\ket}[1]{|{#1}\rangle}
\renewcommand{\braket}[2]{\langle{#1}|{#2}\rangle}
\renewcommand{\Re}{\rm Re}
\renewcommand{\Im}{\rm Im}

\newcommand{\cbe}{\ensuremath{CB_\varepsilon}}

\begin{document}

\preprint{APS/123-QED}

\title{Reduce\&chop: Shallow circuits for deeper problems}

\newcommand{\aqa}{$\langle aQa ^L\rangle $ Applied Quantum Algorithms, Universiteit Leiden}
\newcommand{\lorentz}{Instituut-Lorentz, Universiteit Leiden, Niels Bohrweg 2, 2333 CA Leiden, Netherlands}
\newcommand{\liacs}{LIACS, Universiteit Leiden, Niels Bohrweg 1, 2333 CA Leiden, Netherlands}
\author{Adrián Pérez-Salinas}
\email{perezsalinas@lorentz.leidenuniv.nl}
\affiliation{\aqa}
\affiliation{\lorentz}
\author{Radoica Draškić}
\affiliation{\aqa}
\affiliation{\liacs}
\author{Jordi Tura}
\affiliation{\aqa}
\affiliation{\lorentz}
\author{Vedran Dunjko}
\affiliation{\aqa}
\affiliation{\liacs}
\affiliation{\lorentz}

\begin{abstract}

State-of-the-art quantum computers can only reliably execute circuits with limited qubit numbers and computational depth. This severely reduces the scope of algorithms that can be run. While numerous techniques have been invented to exploit few-qubit devices, corresponding schemes for depth-limited computations are less explored. This work investigates to what extent we can mimic the performance of a deeper quantum computation by repeatedly using a shallower device. We propose a method for this purpose, inspired by Feynman simulation, where a given circuit is chopped in two pieces. The first piece is executed and measured early on, and the second piece is run based on the previous outcome. This method is inefficient if applied in a straightforward manner due to the high number of possible outcomes. To mitigate this issue, we propose a shallow variational circuit, whose purpose is to maintain the complexity of the method within pre-defined tolerable limits, and provide a novel optimisation method to find such circuit. The composition of these components of the methods is called reduce\&chop. As we discuss, this approach works for certain cases of interest. We believe this work may stimulate new research towards exploiting the potential of shallow quantum computers. 

\end{abstract}

\maketitle
\section{Introduction}

Quantum computers have dramatically improved during recent years, with a prominent landmark in the experiments for the so-called quantum computational supremacy \cite{arute2019quantum, zhong2020quantum, madsen2022quantum}.
Despite of these achievements, quantum devices are still far from fault-tolerant computation. A number of limitations is yet to be overcome, including limited number of qubits, decoherence and gate errors. 

Many strategies have been developed to mitigate these issues and reach practical quantum advantages. For instance, error mitigation and supression in general ~\cite{kandala2019error, endo2021hybrid} aims at combating decoherence and noisy gate implementations. The focus on hybrid quantum-classical variational algorithms ~\cite{peruzzo2014variational, farhi2014quantum, cerezo2021variational, bharti2022noisy} aims at finding useful approximations to problems of interest, while minimizing the required quantum resources. Finally, to directly combat size limits and reduce qubit number requirements to accesible quantities, techniques based on circuit cutting have been increasingly studied in literature~\cite{tang2021cutqc, marshall2022high, huembeli2022entanglement, peng2020simulating}. Inspired by qubit-circuit-cutting, we identify a question that remains understudied: can a circuit be cut \textit{in depth}? Or, more generally, to what extent can shallower circuits be used to tackle deeper quantum computations?

In this work, we address the simulation of a deep circuit, given access to many calls to a shallower device. Given a quantum circuit $U$, 
we are interested in computing the quantity $P(x) = |\bra{x} U \ket{0}|^2 $, for a computational basis state specified by the bitstring $x$. We consider ``chopping'' the original circuit into two shallower subcircuits $U = U_2 U_1$, and estimate $P(x)$ by using outcomes from quantum computations relying only on the partial circuits $U_1, U_2$. Our approach is inspired by the so-called Feynman simulation of a quantum circuit~\cite{feynman1995quantum, aaronson2016complexitytheoretic}: the first subcircuit $U_1$ is executed, and the outcomes serve as input for the second subcircuit $U_2$.

The chopping procedure requires consideration of all possible outcomes of the circuit $U_1$, after measuring in the computational basis. We call this number the computational basis rank ($CB$-rank) of the state at the chop. The $CB$-rank dominates the computational complexity of the method. For random states, the $CB$-rank is exponentially large, and a naive implementation of the chopping scheme proposed is not efficient. This motivates the study of scenarios with small $CB$ rank and methods to reduce it. We consider such reduction to be constrained to using a shallow circuit, the {\sl reducer}, for overall savings in depth. This work presents a tailored method to find the reducer using variational strategies. The use of the reducer before chopping the execution gives name to our algorithm: {\sl reduce\&chop}.

The method here exposed presents a series of issues to  overcome for efficient implementation. For arbitrary problems, such issues may be simply unsolvable. For this reason, the present work aims to explore the relationship between deep and shallow circuits, and find interesting cases where depth reduction is possible. 
A faultless recipe to reduce the depth requirements of arbitrary circuits remains out of reach. 

The structure of the paper is as follows:
In~\Cref{sec.reduce-and-chop} we explain reduce\&chop and identify the key bottlenecks.
In \Cref{sec.cb_rank} we investigate the $CB$-rank. In \Cref{sec.theory} we discuss the applicability of reduce\&chop. In \Cref{sec.cb_reduction} we show how to decrease the $CB$-rank and the complexity of reduce\&chop.
In \Cref{sec.numerics} we provide a numerical benchmark for a reduce\&chop algorithm. \Cref{sec.extension} is devoted to possible extensions of the approach. We conclude in~\Cref{sec.conclusions}.

\section{Reduce\&chop for shallower quantum computing}\label{sec.reduce-and-chop}

In this section we describe the reduce\&chop approach by expanding the initial idea of Feynman simulation of a quantum circuit~\cite{feynman1995quantum, aaronson2016complexitytheoretic}. We also identify the arising bottlenecks, and propose methods to mitigate them. Through this work, we assume the regime of noiseless shallow computations and we consider only pure states and unitary operations. 

Our starting point is 
\begin{equation}\label{eq:feynman}
    P(x) = \vert \bra{x} U \ket{0}\vert ^ 2 = \left\vert \sum_{b=0}^{2^n - 1} \bra{x} U_2 \ket{b}\bra{b}U_1\ket 0 \right\vert ^2. 
\end{equation}
The Feynman simulation here used allows us to express $P(x)$ as a combination of shallower computations.  Notably, there is no need to reconstruct the intermediate state $U_1 \ket 0$. Let $d(\cdot)$ denote the depth of a quantum circuit, to be executed in a given hardware setting. The original depth requirement is $d(U)$, but~\Cref{eq:feynman} reduces it to $\max\{d(U_1), d(U_2)\} + c$. The term $c$ is a constant overhead needed to evaluate the relative complex phases between the terms, see~\Cref{sec.amplitudes}. Since $d(U_1) + d(U_2) = d(U)$, the depth requirement is optimized if $d(U_1) = d(U_2)$. 

It is also apparent from~\Cref{eq:feynman} that the cost to evaluate $P(x)$ is dominated by the number of non-zero terms. There can be exponentially many of these terms. We focus on restricting the number of non-negligigble terms of the form $ \bra{b}U_1\ket 0$, enough for obtaining an efficient method. This is equivalent to the number of elements in the computational basis needed to  represent the state $U_1\ket 0$. We dub this quantity the computational basis rank ($\cbe$-rank) of a state. Approximate descriptions with error $\varepsilon$ may also be considered through the $\cbe$-rank.

\begin{definition}[$\cbe$-rank]\label{def:CBd}
Let $\ket{\psi}$ be a quantum state of $n$ qubits, and let $\varepsilon$ be a real non-negative number. The $\varepsilon$-approximate $CB$ rank of a state $\ket\psi$ is 
\begin{equation}
\cbe(\ket\psi) = \min_{K\in\mathbb{N}}\left(K \; {\rm s. t.}\;  \vert\braket{\psi}{\phi_{K}}\vert^2 \geq 1 - \varepsilon\right).
\end{equation}
for some state $\ket{\phi_K}$ with at most $K$ non-zero amplitudes in the computational basis. The value $\vert\braket{\psi}{\phi}\vert^2$ stands for the standard fidelity between two pure states.
\end{definition}

We discuss the general properties of the $\cbe$-rank in~\Cref{sec.cb_rank}. Through this work, $\varepsilon = 0$ if the subscript is missing. The $CB_{(\varepsilon)}$ rank corresponds to the sparsity of the state. However, when dealing with sparse structures one typically assumes sparse access and specification of non-zero elements. Here we do not consider this. 
The idea of $CB$-rank is also closely related to the stabilizer rank in stabilizer-based methods for simulating circuits~\cite{bravyi2019simulation}.

A crucial step in this work is to estimate the $\cbe$ rank of a state. We consider sampling methods for it, which identifythe set $S$ of relevant amplitudes in~\Cref{eq:feynman_reduce2}. The complexity of this procedure grows with the $\cbe$ rank, and thus a reliable and efficient estimation of the $\cbe$ rank is only possible if $\cbe \in \mathcal{O}({\rm poly}(n))$. The details of this statement can be found in~\Cref{sec.estimate_CB}.

\begin{figure}[t!]
    \centering
        \includegraphics[width=\linewidth]{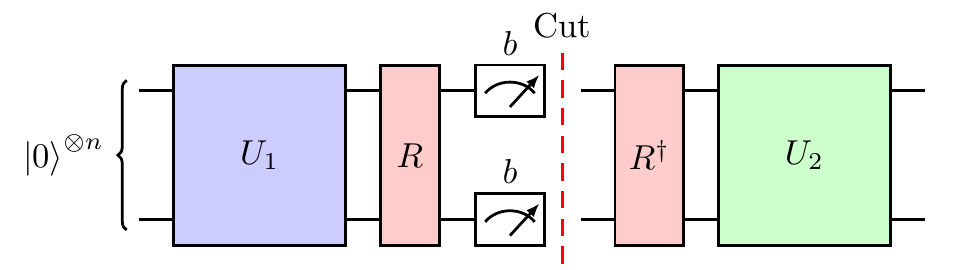}
\caption{Scheme for the sequential splitting of a quantum circuit in two pieces using a single chop. The $U_i$ pieces are determined by the algorithm to run, and variational reducer $R$ is included, applied before the chop and reversed afterwards. The red dashed line stands for the chop. The total depth in each step cannot exceed the hardware limitations.}
    \label{fig.basic}
\end{figure}

Since the complexity of~\Cref{eq:feynman} is dominated by $\cbe(U_1\ket 0)$, we study scenarios with a small $\cbe$ rank, and methods to reduce it. These methods must be shallow to maintain the depth-savings motivating reduce\&chop.
We observe that shallow circuits can transform low $\cbe$-rank states into high $\cbe$-rank states, i. e. $H^{\otimes n}\ket 0^{\otimes n}$. Hence, the converse may also be possible. We aim to find a different circuit $R$, which we call the {\sl reducer}, capable of reducing the $\cbe$-rank, with depth $d(R)\ll d(U_1) $.
Having access to such an $R$, allows us to perform a Feynman simulation as given by  
\begin{align}\label{eq:feynman_reduce}
P(x) = \left\vert \sum_{b=0}^{2^n - 1} \bra{x} U_2 R^\dagger \ket{b}\bra{b} R U_1\ket 0 \right\vert ^2 \\ \label{eq:feynman_reduce2}\approx  
\left\vert \sum_{{ b \in S\subseteq \{0,1\}^{n}, \atop |S| = \cbe}} \bra{x} U_2 R^\dagger \ket{b}\bra{b} R U_1\ket 0 \right\vert ^2 
,
\end{align} 
and it will still lead to a (modest) depth saving.  The depth requirements are in this case $d(R) + d(U_1) + c$. There is a saving in depth as long as this depth is smaller than the original $d(U)$. In addition, the resulting subcircuit $RU_1$ including measurement overhead must fit in the available hardware. We will discuss the theoretical implications of the existence of such a reducer circuit in \Cref{sec.theory}.

The next step is to estimate the $\cbe$ rank of the intermediate state. We consider sampling methods for it, which identifythe set $S$ of relevant amplitudes in~\Cref{eq:feynman_reduce2}. The complexity of this procedure grows with the $\cbe$ rank, and thus a reliable and efficient estimation of the $\cbe$ rank is only possible if $\cbe \in \mathcal{O}({\rm poly}(n))$. The details of this statement can be found in~\Cref{sec.estimate_CB}.

This leads us to the next challenge: finding an appropriate reducer $R$. Inspired by the field of variational algorithms, and related to the approach to quantum variational autoencoders~\cite{romero2017quantum}, we investigate a variational method, where $R(\bm{\theta})$ is a parametrized circuit with fixed architecture. 
The proximal problem is then to solve $ \operatorname{argmin}_{\bm\theta} \  CB_{\varepsilon} (R(\bm\theta) U_1\ket 0 )$. 
Most optimization procedures require the evaluation of the cost function, in this case $CB_{\varepsilon}$. As mentioned, such estimation is not always possible without exponential overhead. Since in general we expect that the $\cbe$ rank is exponential for random states~\cite{Boixo2018characterizing}, we need to carefully design an optimization method to attain $\cbe$-rank reductions. A description of the variational method is detailed in \Cref{sec.cb_reduction}.

\subsection{Potential problems}

We summarize here the list of bottlenecks appearing for conducting the reduce\&chop algorithm explained in the last paragraphs as:
\begin{itemize}
    \item[(B1)] The estimation of the $\cbe$-rank is only efficient under certain conditions, addressed in~\Cref{sec.cb_rank}.
    \item[(B2)] The existence of a reducer $R$ capable of transforming a given quantum state into a low-$\cbe$-rank state is not guaranteed, addressed in~\Cref{sec.theory}.
    \item[(B3)] The $\cbe$-ranks for states generated by parameterized quantum circuits with random parameters are likely large and non-estimable, addressed in~\Cref{sec.cb_reduction}.
    \item[(B4)] The optimization of the parameters of $R(\bm\theta)$ suffers from several difficulties, addressed in~\Cref{sec.cb_reduction}. 
\end{itemize}

We note that our approach for reducing depth is not the only possible. For instance, we can reduce any arbitrary $d(U_1)$, when applied to $\ket 0$, to $\mathcal O(n)$ by introducing exponentially many ancillary qubits~\cite{sun2021asymptotically}. Reduction to constant depth is also possible, e.g., by using measurement-based quantum computation and (exponentially-costly) post-selection~\cite{briegel2009measurementbased}. Certain versions of the swap test can also be used~\cite{jeffery}.
All the aforementioned alternatives come with a substantial cost in the number of additional qubits needed, and here we focus on methods which require very few ancillas. It is also possible to generalize the reduce\&chop approach to many cuts, see~\Cref{app:feynman}.

Finally, we emphasize that the underlying objective of this work is not to provide effective algorithms which can be employed on current devices to run deep circuits reliably. Rather, we use this approach to probe the relationship between deep and shallow computations, especially in a variational-algorithm context, and to provide some guidelines on how shallow circuits could be used for deeper problems. 

\section{($CB$)-rank and characterization of state at the chop}\label{sec.cb_rank}
In this section we review the basic properties of the $\cbe$-rank, and provide the method to evaluate it. This  section also addresses bottleneck (B1). 

We recall the formal statement of $\cbe$ rank of a given state from~\Cref{def:CBd}.
The immediate question is how to approximate a given state with another low-CB-rank one. For a state $\ket\psi$, it is straightforward to find $\ket{\psi_K},$ with  $CB(\ket{\psi_K}) = K$ with that maximizes $\vert\braket{\psi}{\psi_K}\vert$. It suffices to choose the $K$ largest coefficients and discard everything else, with proper normalization.
\begin{lemma}\label{approx_cb_rank_state}
    Consider a quantum state $\ket{\psi} = \sum_{i = 0}^{N - 1}\alpha_{\pi(i)}\ket{\pi(i)}$ where $\pi$ is a permutation such that $|\alpha_{\pi(i)}|^2 \geq |\alpha_{\pi(i + 1)}|^2$ for all $i \in \{0, 1, \dots, 2^n - 2\}$. The maximum absolute overlap between  $\ket{\psi}$ and any state with at most $\cbe$-rank $K$ is $\sqrt{1 - \sum_{i = 0}^{K - 1}|\alpha_{\pi(i)}|^2}$ and it is achieved by the state 
    $$\ket{\psi_{K}} \propto \sum_{i = 0}^{K - 1}\alpha_{\pi(i)}\ket{\pi(i)}, $$
    when normalized. 
\end{lemma}
The proof can be found in~\Cref{app:proof}.

In general, the $\cbe$ rank can be exponentially large, even in the case of the output of extremely simple circuits applied to a state with $CB$-rank 1. An example is a single layer of Hadamard operations acting on the $\ket 0$ state. However, output states of more complex circuits can have much smaller $CB$ ranks. For example, one needs $n$ operations and $n$ depth to generate the GHZ state from $\ket 0$, with $CB = 2$, using at most two-qubit gates, i.e. using a cascade of CNOTs. 

We consider here some geometrical properties about sets of states with $CB$-rank $K$. Clearly, for $K=1$ these form a finite set of $2^n$ elements, one for each computational basis vector. Alternatively, one can think of it as the union of $2^n$ one-dimensional rays if the normalization condition is dropped. For other $K$'s, this set is the union of ${2^n \choose K}$ $K$-dimensional subspaces, each spanned by the choice of $K$ computational basis vectors, and there are ${2^n \choose K}$ possible choices. Hence, for $K<2^n$,$\mathcal K$ forms a zero-measure subset. For $K>1$, $\mathcal K$ forms a single connected component. In extending the notion to $\cbe$-rank, these have non-zero measure for all $K$.

A motivation for developing reduce\&chop is to interrupt a quantum circuit before significant noise affects the state, thus we focus on the simple case with no noise, in this first analysis. 
However, we can briefly discuss the effects of small amounts of noise. For simplicity, we will avoid explicit models of noise and consider the noise-channel to depend on depth only, and fix any other parameters. We assume the error is upper bounded by $\delta(d)$ with respect to a distance, e.g. trace distance, between noisy and zero-error states. Each circuit half in the reduce\&chop procedure accumulates error up to $\delta(d(U_{1, 2}) + c)$. For $U_2$, many executions with different input state and same error bounds are necessary. However, the same bound in the total error holds due to the small contribution of each input state to the final computation $p(x)$.  The estimation of the state at the chop is bounded to error $\epsilon$. Assuming $d(U_{1} = d(U_{2})$, the reduce\&chop procedure is advantageous from an error perspective if $\epsilon < \delta(d(U)) - 2 \delta(d(U) / 2 + c) $, that is if noise is superadditive in depth.

Simple noise models, e.g. exponentially decaying thermal relaxation, do not have this property. However, reduce\&chop still suits in cases, such as practical implementations where the depth is upper bounded, e. g. optical arrays; or noise channels with increasing-in-depth decay rate, such as dynamical distortion in superconducting qubits~\cite{rol2020timedomain}.

\subsection{Estimation of $\cbe$-rank}\label{sec.estimate_CB}
The estimation of $\cbe$-rank from measurements needs a measurement budget $M \in \Omega(CB_{\varepsilon})$. For any state with exponential $\cbe$ rank, such as the output of random circuits~\cite{Boixo2018characterizing}, the exponential scaling of the required number of measurements prevents any implementation of reduce\&chop without exponential classical overhead. 

We describe here a procedure to estimate the value of the $\cbe$-rank of a quantum state by preparing and measuring the state $M$ times. 
The procedure is given in detail in~\Cref{alg:measurement}. In summary, this procedure samples twice from the quantum state of interest $\ket\psi$, to return a candidate $K$ for the $\cbe$-rank, and an upper bound in ${\rm Prob}(K < \cbe)$, accepted only if sufficiently small. The two samplings are needed to define a Bernoulli process required to formulate the following~\Cref{le.CB^{(M)}easure}, which gives support to this procedure.

\begin{algorithm}[t!]
\SetKwInOut{Input}{Input}
\SetKwInOut{Output}{Output}
\SetInd{.25em}{.5em}
\caption{Algorithm to estimate $\cbe$-rank}\label{alg:measurement}
\Input{Quantum circuit generating $\ket\psi$, the quantum state whose $\cbe$-rank we want to estimate}
\SetKwInOut{Input}{}
\Input{$M$, the number of measurements available}
\Input{$\varepsilon$, the precision threshold to compute the $\cbe$-rank}
\Input{$p_m$, the minimum failure probability accepted}
\Output{$K$, estimation of the $\cbe$-rank}\SetKwInOut{Output}{}
\Output{$p$, probability of underestimation of the $\cbe$-rank}
\Output{$F$, flag for success of the algorithm. }
Sample bistrings $i$ from the quantum state $\ket\psi$, in two sets, each with $M$ samples\;
$\{(x_i, n_i, m_i)\}_i \gets $ Bitstring $(x_i))$ and its number of outcomes after $M$ measurements, for each sampling $a = \{1, 2\}$\;
$\{(x_i, n_i, m_i)\}_i \gets $ Order $\{(x^{(a)}_i, n_i, m_i)\}$ in descending order in $n_i$\;
$p \gets 1$\;
$m \gets M$ \;
$C \gets \vert \{x^{(1)}_i\}\vert$, number of bitstrings found during the first sampling\; 
\For{$i \gets 1$ \KwTo $C$}{
$m \gets m - m_i$\;
\If{$m < M\varepsilon$}{
$p \gets {\rm exp}\left(-2M\left(\varepsilon - \frac{m}{M}\right)^2 \right)$\;
\If{$p < p_m$}{
$K \gets i$\;
$F \gets {\rm True}$\;
\Return{$K$, $p$}  
End algorithm\;
}
}
}
$K \gets C$\;
$F \gets {\rm False}$\;
\Return{$K$, $p$}\;
\end{algorithm}

\begin{lemma}\label{le.CB^{(M)}easure}
Let $\ket\psi$ be a quantum state. This state is prepared and measured $2M$ times, and the outcomes separated in 2 sets of $M$ samples each. Let $K$ be any integer selecting $X_K$, the $K$ most frequent outcomes in the first sampling. Let $m$ be the number of outcomes in the second sampling with outcomes not in $X_K$. If $m < M\varepsilon$, then 
\begin{equation}
{\rm Prob}\left(K < \cbe\right) \leq \exp\left(-2M\left(\varepsilon - \frac{m}{M} \right)^2\right).
\end{equation} 
A proof is detailed in~\Cref{ap:cb_rank}.

\end{lemma}
This result is a direct application of Hoeffding's inequality~\cite{hoeffding1963probability} to Bernoulli probability processes.

The range of applicability of~\Cref{alg:measurement} depends on the available measurement budget $M$. In this work, we consider the regime $\cbe \leq CB^{(M)}$, and any $\cbe$-rank must be reliably estimable. We set $M = CB^{(M)} \varepsilon^{-2}$ to ensure that the sum of all discarded outcomes is estimated within the desired confidence interval. However, depending on the underlying probability distribution we may be able to reliable estimate larger $\cbe$ ranks.

\subsection{Estimation of amplitudes}\label{sec.amplitudes}

\begin{figure}[h!]
\centering
\subfloat[Hadamard test with one ancilla]{\includegraphics[width=.45\linewidth]{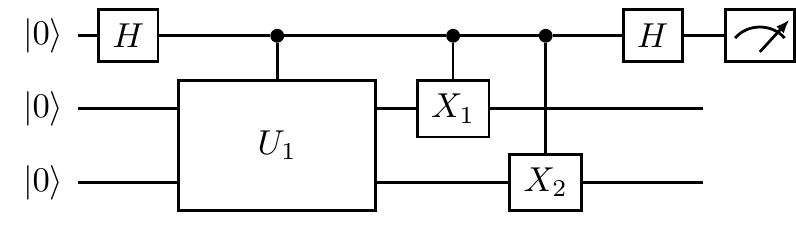}
\label{fig.hadamard1}}
\subfloat[Hadamard test with $n$ ancillas]{
\includegraphics[width=.45\linewidth]{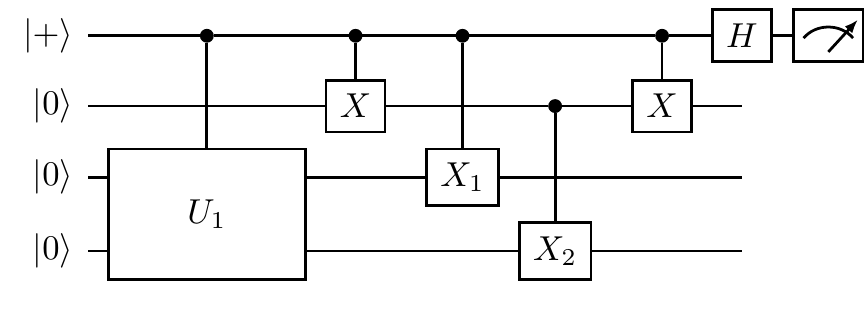}
\label{fig.hadamard2}}
\caption{Schemes of the circuits needed to measure the overlap between the state $U_1\ket 0$ and $\ket x = \cdots X_2 X_1 \ket 0$, where $\cdots X_2 X_1$ summarizes all operations required to obtain the bitstring $x$ from $\ket 0$. To compute real and imaginary parts, it suffices with adding a convenient phase gate before the measurement.}\label{fig.measures}
\end{figure}

In order to estimate $P(x)$ from~\Cref{eq:feynman}, we need to estimate the amplitudes $\bra{b} U_1\ket{0}$. This can be done with Hadamard tests~\cite{aharonov2006polynomial}. The elements $b$ with large contributions to $P(x)$ are identified via the estimation of $\cbe$-rank described in~\Cref{alg:measurement}. The Hadamard test can be implemented using just a single ancillary qubit in the $\ket +$ state, requiring $n$ extra layers, or using a $n$-qubit GHZ states and a swap network, with overhead of only $1$ layer in depth. This method allows us to estimate real and imaginary parts of the elements factors in~\Cref{eq:feynman}, and thus also relative complex phases. See~\Cref{fig.hadamard1} and~\ref{fig.hadamard2} for a description.

\subsection{Error estimation}

The methods here described allow to obtain an estimation of the state at the chop within a certain accuracy that depends both on  $\varepsilon$, the estimated $\cbe$-rank $K$ and the number of measurements. An estimation of the errors establishes the following result.

\begin{lemma}[Error estimation]\label{le.error}
Let $\ket\psi$ be a quantum state, and let $\ket{\hat\psi_K}$ be its approximation after chop with measurement error, with $\cbe$-rank $K$. The output probability of each bitstring is estimated using the procedure described in~\Cref{alg:measurement} using $M_p$ samples, with $m$ outcomes outside the $K$ most relevant ones. The real and imaginary parts of each coefficient are estimated using Hadamard test with $M_\varphi$ measurements. Then 
\begin{equation}
    \vert \braket{\psi}{\psi_K}\vert^2 \geq  1 - \varepsilon - \frac{K}{2 M_\varphi (1 - m/M_p)} 
\end{equation}
with probability at least $1 - \exp\left(-2M_p(\varepsilon - m / M_p)^2\right)$.
\end{lemma}
A detailed calculation of this error can be found in~\Cref{app:estimation_error}.

\section{Theoretical underpinnings}\label{sec.theory}
In this section we briefly discuss the theoretical underpinnings and implications of the proposed method. We also address the (B2).

\subsection{Considerations on complexity of the method}

The idea that deep circuits could be simulated with shallower circuits efficiently is connected to the Jozsa conjecture~\cite{jozsa2005introduction}: any BQP computation can be evaluated using polynomially many calls to log-depth quantum computations, assisted with overall polynomial classical processing. If Jozsa conjecture is true, then the circuit is classically simulatable. This conjecture is now known not to hold relative to oracles \cite{coudron2020computations} but still may be true in other scenarios, such as the ones we consider in this work. 
For our approach to match the Jozsa conjecture, we would require $D/\log(D)$ cuts, for $D \in {\rm poly}(n)$. This would incur a cost $k^{D/\log(D)}$, where $k$ constitutes the query complexity of just one cut. For an overall polynomial complexity, $k$ needs to converge to unity as circuits grow. Our approach is more pragmatic and tackles more modest, non-asymptotic, depth savings. We focus on achieving simulation with constant depth-reductions and polynomial overheads. 

Another interesting consideration is the $\cbe$-rank of (the output state of) a circuit at an arbitrary cut. 
If any cut in a given circuit constitutes a low $\cbe$ state, then it is easy to see that circuit is classically simulatable via matrix product state (MPS) methods~\cite{vidal2004efficient}. However, in our case we only require the cut accross one point in the circuit to have low $\cbe$-rank. Each subcircuit $U_1, U_2$ can be arbitrarily hard to simulate. This constitues a substantial difference between our approach and MPS. 

Our method critically requires sampling from the circuit $U_1$, even if we are interested in computing expectation values at the end. Sampling  sampling from shallow circuits is likely computationally harder than computing expectation values~\cite{terhal2004adaptive, Bravyi2021classical, franca2021limitations,Boixo2018characterizing}. This has been recently questioned for random circuit sampling in the noisy regime~\cite{aharonov2022polynomialtime}, but not in general. This raises confidence that our approach does not likely lead to easy-to-classically-simulate methods. 

\subsection{Existence of shallow reducers}

Let us first discuss the reasons why shallower-depth reducers without ancillas should exist. To this end, consider the problem of computation of $\left\vert \bra{x} U_2 U_1\ket 0 \right\vert ^2$. We assume  $U_1$ and $U_2$ are given by circuits of lowest possible depth\footnote{With respect to a given gate set.}, and that the $CB_{\varepsilon}(U_1\ket 0)$-rank is large. For simplicity $\varepsilon$ is small, but not specified.
Does there exist a unitary $R$ such that $\cbe(RU_1 \ket 0)$ is low, while $d(R) \ll d(U_1)$?

First, we make a few observations. 
\begin{inparaenum}[(1)]
\item There exists an $R$ which exactly, and maximally reduces the rank, at the cost of $d(R)  = d(U_1)$: $R = U_1^\dagger$. 
	\item It would be enough to require $\ket 0=RU_1 \ket 0$, that is, $R$ only needs to invert the one-dimensional input space $\ket 0$. 
    \item It is sufficient for $R$ to transform $U_1\ket 0$ into a state of poly-sized $CB$-rank, rather than $1$.
    \item We do not require the exact ($\varepsilon=0$) $CB$-rank to be low: just the $\cbe$-rank for finite $\varepsilon$ values.
\end{inparaenum}

Each of the above observations opens the possibility for a lower-depth $R$ by reducing the constraints on $R$.
While intuitive, the observations above do not yield quantitative guarantees about the existence of a shallower reducer. This can be improved by stating more precise statements. 
Define the circuit distance between two states $\ket{\psi}, \ket{\phi}$ as \footnote{The equality can be softened to approximate equality.}. 
\begin{equation}
D_c(\ket\psi, \ket\phi) = \min_W \left(d(W) \ {\rm s. t. } \ \ket\psi = W \ket\phi \right),
\end{equation}
for $W$ being a circuit constructed from a given gate set, and $d(W)$ is its depth.
The circuit distance between a state $\ket{\psi}$ and a set of states $S$ is the the minimal distance between $\ket{\psi}$ and any state in $S$.
\begin{equation}
D_c(\ket\psi, S) = \min_{\ket\phi \in S} D_c(\ket\psi, \ket\phi)
\end{equation}
If we set $\ket{\psi} =U\ket 0$, in our approach there exists a shallow reducer if
\begin{align}
D_c(U \ket 0, \mathcal K) & \ll D_c(U \ket 0, \ket 0) \\
\mathcal K & = \left\{\ket\psi \;{\rm s. t. }\; CB(\ket\psi) \leq K \right\}.
\end{align} 
The $\ll$ symbol aims to emphasize the shallow nature of the reducer. 
Remember that we impose $K \in \mathcal O( {\rm poly}(n))$ to guarantee efficiency. Since $\ket 0 \in \mathcal{K}$, the condition $D_c(U \ket 0, \mathcal K) \leq D_c(U \ket 0, \ket 0)$ is guaranteed. Although it is not clear whether and when there exist states $\ket\psi$ for which the distances to $\ket 0$ and $\mathcal K$ are in fact the same, it is easy to see there are states for which these are very different. For example, if $\ket\psi \in \mathcal K$, by definition $D_c(\ket\psi, \mathcal K) = 0$, irrespective of $D_c(\ket\psi, \ket 0)$. 
The above class generalizes easily: for example, let $\ket{\psi} \in \mathcal K$ require depth $\Omega(n)$ to be prepared; then if $Q$ is a constant depth circuit, in general $Q\ket{\psi} \notin \mathcal K$, yet it still remains at $\Omega(n)$ distance from $\ket 0$ but constant distance from $\mathcal K$\footnote{Proof: by contradiction. Let $P$ be a sub-linear depth circuit s.t. P $\ket{0} =  Q\ket{\psi}$; then $Q^\dagger P \ket 0 = \ket \psi $, and $ Q^\dagger P$ is obviously sub-linear depth.}. 

This suggests that indeed $R$ could be significantly shallower than $U_1$, as, intuitively, we can delegate the (costly) preparation of a good initial low-$CB$ state to classical processing. Only the amplitudes and the corresponding element in the computational basis must be stored, with no additional overhead for a particular choice of the elements needed for representation. 
Recall that preparing sparse state with polynomial $CB$-rank from a basis state requires polynomially many gates, and thus, in general, polynomial depth. 
Polynomially many ancillary qubits and unitaries acting on larger numbers of qubits can relax this depth requirement~\cite{zhang2021lowdepth}.

\subsection{Finding reducers}

Next, we turn our attention to the more difficult problem of finding a suitable $R$. In our scheme we utilize a two-stage optimization process, to be detailed in~\Cref{sec.cb_reduction}, where we consider parametrized $R(\bm\theta)$ and $U_1$. This motivates us to study the $CB$-rank properties of the set of states $R({\bm \theta})U_1 \ket{0}$, for valid instances of $U_1$.
In this discussion, we shall denote ``low" and ``high" $CB$-rank states as those below and above a desired threshold. 

We define first different sets of states related to the operators $U_1 \in \mathbb U \subseteq \mathcal{SU}(N)$,  and $R(\bm\theta)$. First, we define
\begin{equation}
    \mathcal{U} = \lbrace \ket \psi = U_1\ket{0^n}, \quad \forall U_1 \in \mathbb U \rbrace,
\end{equation}
Notice that this set is entirely dependent on the subset $\mathbb U$. We also define a set of states that are $\varepsilon$-close to $\mathcal{U}$ as
\begin{equation}
\mathcal{U}_\varepsilon = \lbrace \ket{\psi_\varepsilon} \textrm{ s.t. } \; \exists\; \ket\psi \in \mathcal U \, : \, \vert \braket{\psi_\varepsilon}{\psi} \vert^2 \geq 1 - \varepsilon \rbrace.
\end{equation}
We define the set of states that can be reduced by $R(\bm\theta)$ to another low-$CB$ state, in particular with $CB_{0}$-rank upper bounded by $K$, as
\begin{equation}
    \mathcal{CB}_K = \left\lbrace \ket \psi = R^\dagger({\bm \theta}) \ket{K}, \quad {\rm with} \, \ket{K} \in \mathcal K \right\rbrace,
\end{equation}
where $\pi(i)$ is any permutation of the computational basis. Recall that $\mathcal K$ is the set of states with $CB$ rank at most $K$. 

\begin{figure}[t!]
    \centering
    \includegraphics[width=.8\linewidth]{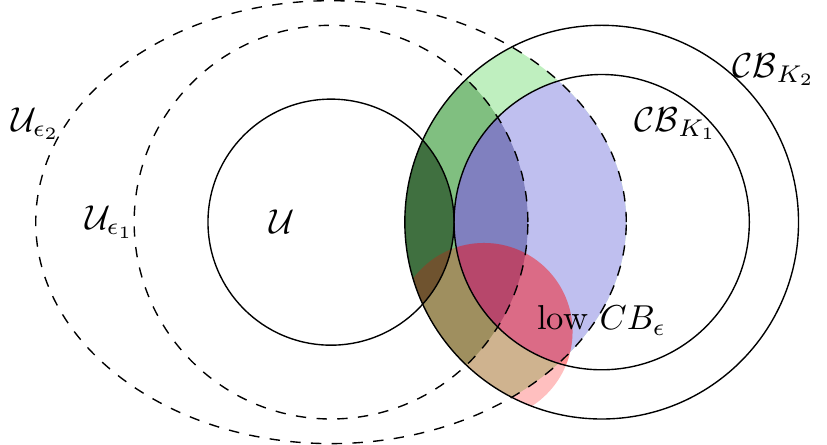}
    \caption{Schematic representation of the distribution of the quantum states with respect to its $CB$-rank reduction. 
The depth-reduction algorithm can only be applied to those states in the intersection of both sets. The low $CB$ area corresponds to those states with small $CB$ before applying $R$. Several possible configurations may arise in different cases, and the exact relationships between state sets depend entirely on the circuits used to prepare them. The sets are explicitly defined as:
   $\mathcal{U}$) Set of states generated by the circuit to be chopped $U_1({\bm \phi})$;
    $\mathcal{U}_\varepsilon$) Set of states $\varepsilon$-close to $\mathcal{U}$, with $\varepsilon_2 > \varepsilon_1$;
    $\mathcal{CB}_K$) Set of states that can be reduced to another state with $CB$ rank below $K$ using a reducer $R({\bm\theta})$, with $K_2 > K_1$.
    }
    \label{fig.sets}
\end{figure}

The circuits whose depth can be reduced using the reduce\&chop algorithm are those in the intersection of the sets defined above, that is
\begin{multline}
    \mathcal{U}_\varepsilon \cap \mathcal{CB}_K = \\ \lbrace \ket\psi = U_1\ket{0^n} \mbox{ s.t. } \cbe(R({\bm \theta}) \ket\psi) \leq K \rbrace.
\end{multline}
Among the states $\ket\psi \in \mathcal{U}_\varepsilon \cap \mathcal{CB}_K$ it is possible to distinguish two subsets, namely those with high and low $\cbe$-rank before the application of the reducer $R$. For low $\cbe$ states the reducer is trivial. In the case of high $\cbe$ states, the key challenge is to find the optimal values of $\bm\theta$ for the reducer. 

It is worth to further explore the role of $\varepsilon$. Consider the set of states generated by the reducer applied to some initial state $\ket\psi$, $\mathcal R_{\bm\theta, \ket\psi} = \left\lbrace R(\bm\theta)\ket\psi \right\rbrace_{\bm\theta}$. This is the set of states we can access during the optimization, which may contain states of different $CB$-ranks. The optimization process aims to always find those states with smaller $CB$-ranks. Considering approximate $\cbe$-ranks is equivalent to extending the previous set to $\mathcal R_{\bm\theta, \ket\psi, \varepsilon} = \bigcup_{\bm\theta}\mathcal B_\varepsilon\left( R(\bm\theta)\ket\psi \right)$, where $\mathcal B_\varepsilon\left( \ket\phi \right)$ denotes the set of states with fidelity larger than  $1 - \varepsilon$ with respect to $\ket\phi$. This set uniformly grows in $\varepsilon$ and it becomes more likely to obtain states with smaller $CB$-ranks. In our optimization algorithm 
we explore the space parametrized by $\bm\theta$, and additionaly access the $\mathcal B_\varepsilon$-regions around the $\bm\theta$ points. 

The system size plays a role in the hardness of finding reducers using a brute force approach. However, heuristic methods, such as the one detailed in~\Cref{sec.cb_reduction} might help for an efficient search of such reducers.

\section{Variational method for $\cbe$-rank reduction}\label{sec.cb_reduction}

The theoretical underpinnings described in~\Cref{sec.theory} discuss the existence of shallow $R(\bm\theta)$ capable of reducing the $\cbe$-rank of a given state. However, to the best of our knowledge there is no straightforward recipe to find it. This section addresses the obstacles arising in the path towards successful optimization. We search for the optimal values of $\bm\theta^\star$ as
\begin{equation}
    {\bm \theta}^\star = {\rm argmin}_{\bm\theta} \left( \cbe(R({\bm \theta})U_1\ket 0) \right),
\end{equation} 
where $\varepsilon$ is fixed and treated as a hyperparameter. $R$ will favor a concentration of the amplitudes into a few computational basis states. The optimization of the $\cbe$-rank is however far from trivial. First, the $\cbe$ function is an integer, discontinuous and locally flat. In addition, the optimization is involved as the objective function is non-convex.

\begin{figure}
\centering
\includegraphics[width=\linewidth]{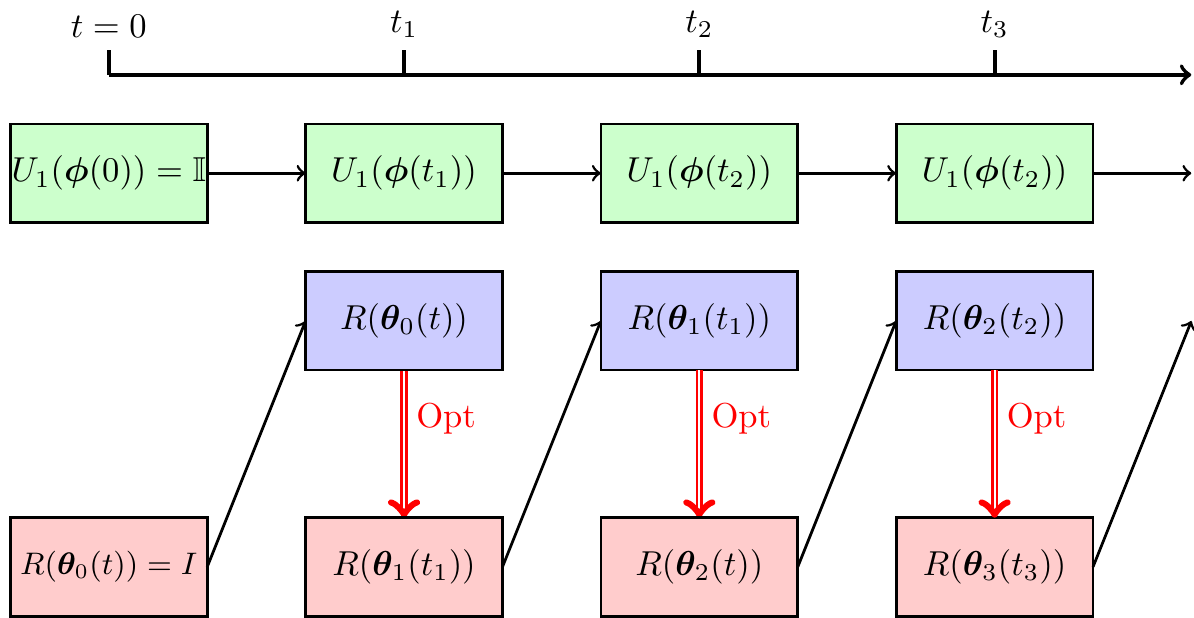}
\caption{Gradually activated optimization algorithm for finding the optimal values of the reducer $R$ defined using variational circuits. The times $t_i$ are reached when the $\cbe$ rank exceeds a certain threshold $CB^{(M)}$. }
\label{fig.grad_opt}
\end{figure}

We consider the case where we have an allowed budget of measurements to estimate the $\cbe$-rank. Considering conventional notions of efficienty, for $M$ in $\mathcal{O}\left({\rm poly}(n)\right)$ only polynomial values of $\cbe$-rank can be reliably estimated. This is bottleneck (B3). 
In order to keep the values of the $\cbe$-rank always under a certain threshold, a two-tier approach has been designed, summarized in~\Cref{fig.grad_opt}.

First, we note that common inputs to our problem are in practice circuit specifications $U_1(\bm\phi)$, rather than black-box unitaries. We propose a gradual activation of $U_1$, meaning that the parameters $\bm\phi$ follow a path, defined by an activation parameter $t\in[0,1]$. Such path satisfies $U_1({\bm\phi}(0)) = {\mathbb I}$ and $U_1({\bm\phi}(1))=U_1$. 

For the optimization process, we activate $U_1$ by increasing $t$. This step stopswhen the $\cbe$-rank reaches the stopping threshold $CB^{(M)}$. The reducer $R(\bm\theta)$ is then optimized. After optimization is complete, the increment of $t$ is continued. 
This recipe aims to find a path in the parameter space $\bm\phi \cup \bm\theta$ which avoids regions of high $\cbe$ rank at the cut. The existence of such path is assumed, and numerically observed, although no guarantees exist. 
The gradual activation of $U_1$ may also generate states of the form $\ket\psi \propto \ket{{\rm low}\ \cbe} + \delta \ket{{\rm high}\ \cbe}$, with $\delta \leq \varepsilon$. If $\delta$ increases as $U_1$ is activated, there is a discontinuity in the $\cbe$ rank for $\delta = \varepsilon$. If such change is too large, the optimizer may never see the gradual activation of $U_1$. 
On top of these general comments, the explicit activation path has a relevant role in the optimization process, and must be chosen for each instance of the problem. 

For each step in the optimization approach, we need an optimizer capable to reduce $\cbe(R(\bm\theta)U_1({\bm\phi}(t)\ket 0)$ with respect to $\bm\theta$. As the first ingredient, we propose the cost function for the optimizer to be
\begin{equation}\label{eq.loss_function}
\mathcal{L}({\bm \theta}, t) = K_\varepsilon({\bm \theta}, t) - \log(1 - p_\varepsilon(\bm \theta, t)), 
\end{equation}
where $(K_\varepsilon(\bm \theta, t), p_\varepsilon(\bm \theta, t))$ are the output of~\Cref{alg:measurement}. The loss function consists of two pieces: The estimation $K$ corresponds to the $\cbe$-rank to minimize. The second term in~\Cref{eq.loss_function} exploits the $\varepsilon$-robustness of the $\cbe$ rank to make the loss function continuous. Although the discrete nature of the $\cbe$-rank does not prevent the optimizer to explore the landscape, we observed this loss function to be useful for avoiding premature convergence. It also prevents the optimizer to explore areas of the landscape where the $\cbe$-rank cannot be reliably estimated, by artificially increasing $\mathcal L$. For general architectures and random $\bm\theta$, this loss function diverges~\cite{Boixo2018characterizing}. 

Now we need the method for optimization. One could consider gradient-based methods to optimize the loss function. However, numerical gradients are a non-feasible approach (B4) from a practical perspective, especially in noisy settings. Standard gradient shift rules are not applicable because \begin{inparaenum}[(1)]
\item the loss function is not constructed as an expectation value, and
\item it is possible to construct circuits whose $\cbe$ blows up for small parameter changes, preventing the efficient estimation of its value.
\end{inparaenum}
Thus, standard methods are not applicable~\cite{crooks2019gradients, wierichs2022general}.

We propose evolutionary strategies (ES) as a solution to (B4). ES can find a compromise between landscape exploration and staying in the neighbourhood of a point with estimable $\cbe$. ES sample points in the landscape from an probability distribution, updated in each iteration. Ideally, the probability of finding low values of the loss function increases with each iteration. The choice of ES rely on the following observation. These strategies explore the neighbourhood of a point in the paramter space. As noted before, even a small change in the parameters can make the $\cbe$ increase drastically and prevent us to compute it efficiently. Gradient methods do not work for this reason. However, since an ES probabilistically explores the parameter space, we assume the method to find at least some points where the $\cbe$ is estimable.

We observed during numerical experiments the importance of the tunable spread $\sigma$ in the sampling of new points. Local exploration with a small $\sigma$ usually delivers states with small changes in the $\cbe$-rank (at least for those states with error bounded away from $\varepsilon$). As a trade-off, the effective step size in the search space is small, and the optimization process requires more iterations. In turn, large $\sigma$ allows for a rapid exploration. However, this setting implies a much higher risk of landing in states whose $\cbe$-rank is too large to be estimated efficiently. for our problem need a sweetspot between these to scenarios. 

All the assumptions for ES can be summarized as \begin{inparaenum}[(1)] \item the existence of a path in $\bm\theta$ such that the $\cbe$ rank is estimable and \item the capability of the ES to explore this path. \end{inparaenum} The path does not have to be continuous, as long as at least one of the points reached by the ES in each iteration returns satisfactory results. If this is the case, (B1) is avoided. 

As stated in~\Cref{sec.estimate_CB}, there is small a failure probability in the estimation of the $\cbe$-rank. This failure probability depends on individual steps, and not on the optimization path. If the optimization is successful, this failure probability would lead, in the worst case, to a slight underestimation of the $\cbe$ rank with no further consequences. 

Finally, we mention that everything we discussed so far is heuristic and relies on reasonable but difficult-to-check assumptions,
independently from the method. To analyze the goodness of our proposals, and the performance of this approach we conducted numerical experiments in~\Cref{sec.numerics}. 

\section{Numerical experiments}\label{sec.numerics}

In this section we experimentally investigate the reduce\&chop method for a particular illustrative example and study its numerical behaviour. The simulations are available in Ref.~\cite{code}, and the framework to simulate quantum circuits was {\tt qibo}~\cite{efthymiou2022qibo}.

\paragraph{Problem and Ansätze:}The problem we consider is to mimic the execution of a deeper circuit by chopping it in two halves as in the reduce\&chop algorithm previously explained. 

\begin{figure}[t!]
    \centering
    \subfloat[TFIM-inspired Ansatz, $U$]{
        \includegraphics[width=.5\linewidth]{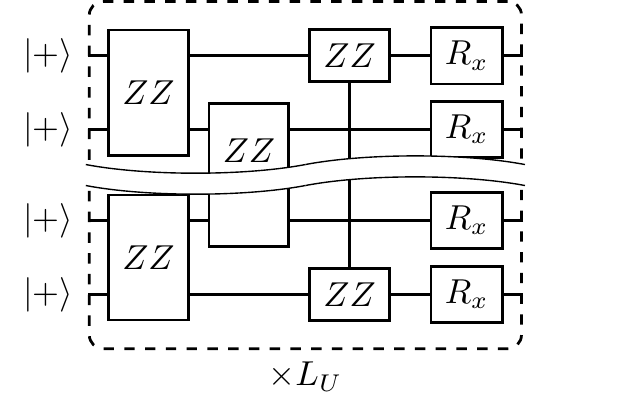}
        \label{fig.tfim_Ansatz}}
    \subfloat[HEA Ansatz, $R$]{
       \includegraphics[width=.4\linewidth]{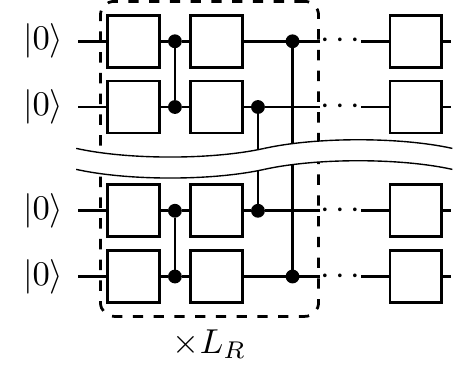}\label{fig.hea_Ansatz}}
    \caption{Circuits for the considered input circuit to be chopped with the depth-limited quantum computing algorithm. The blank gates are arbitrary single-qubit gates, and for labeled gates $R_x = e^{-i \sigma_x \theta / 2}$ and $ZZ = e^{-i \sigma_z \otimes \sigma_z \theta / 2}$. The wavy line stands for many qubits not explicitly drawn in the graphics.}
    \label{fig.Ansatz}
\end{figure}

We choose $U$ to be Transverse Field Ising Model (TFIM) Hamiltonian-inspired Ansatz, see~\Cref{fig.tfim_Ansatz}. The motivation for this choice is that this kind of Hamiltonian-Inspired Ansatz explores the Hilbert space slower than a generally expressive Ansatz (e.g. a $2$-design), increasing the chance of finding an optimal reducer. 

The reducer $R$ follows a layered Hardware-Efficient-Ansatz (HEA) which intersperses single-qubit arbitrary operations and entangling CZ gates. The entangling gates are laid out in a cyclic graph alternatively connecting neighbor qubits. 
Additionally, the final step is an extra set of single-qubit operations. See~\Cref{fig.hea_Ansatz} for the circuit. 
The motivation of this choice is to generate an Ansatz with high expressivity and shallow depth, so that its reduction capabilities reach many states. Sure this kind of Ansatz can encounter the problem of barren plateaus~\cite{mcclean2018barren}, in particular for deep reducers. Addressing barren plateaus in this context is out of the scope of this work, and we still consider the in-principle feasibility of reduce\&chop worthy. 

The parameters for the circuits are randomly drawn from a uniform distribution. The depth is computed in terms of how many entangling layers are implemented in the circuit in sequential steps. In the TFIM case, the depth is $4 \times L_U$, since each $ZZ$ gates requires two CNOTs. For the HEA Ansatz, the depth is $2\times L_R$. 

\paragraph{Optimizer:} The optimizer chosen for the experiments here conducted is CMA~\cite{hansen2006cma}, implemented using the default options in {\tt CMA-package}~\cite{niko2020cmaes}. The only custom choice is the spread hyperparameter $\sigma$.
This covariance is set to $\sigma = \varepsilon$. With this choice we prevent the optimizer from exploring areas of the landscape with high $\cbe$-ranks.

\paragraph{Hyperparameters:} 
The reduce\&chop algorithm has been tested for 8 and 10 qubits, and $U$ with depth $40$, $R$ with depth $4$. The stopping criteria for $CB^{(M)}$ is set to $n^3 / 5$, and the measurements to estimate the $\cbe$-rank and perform the Hadamard tests are both $\varepsilon^{-2}\,n^3 / 4 $. With this option, the maximum estimable $\cbe$-rank is above the stopping criteria. There is a loose margin for the optimizer to evolve without reaching non-estimable values of the $\cbe$-rank. The state after the chop is reconstructed with the information of the measurements. The failure probability is set to $p_m < 10^{-4}$. The failure probability is monitored during the process to ensure its value remains below the threshold $p_m$ throughout. Finally, the considered values for the error are $\varepsilon = \{0.02, 0.03, 0.05, 0.08, 0.13 \}$. 
All these parameters are chosen according to the available computational capabilities. 

\paragraph{Gradual activation schemes:} We consider two different activation schemes, namely the soft activation and the parametric activation. In the soft activation, the parameter $t$ acts as $\ket{\psi(t)} = \left(\cos(\pi t / 2)\mathbb{I} + \sin(\pi t / 2) U_1(\bm\phi)\right) \ket 0$. In the case of the parametric activation, the different parameters $\bm\phi$ of the Ansatz $U_1(\bm\phi)$ are activated sequentially one by one.

\subsection{Results}

\begin{figure}
\centering
\subfloat[8 qubits, $\varepsilon = 0.02$]{
\includegraphics[width=.45\linewidth]{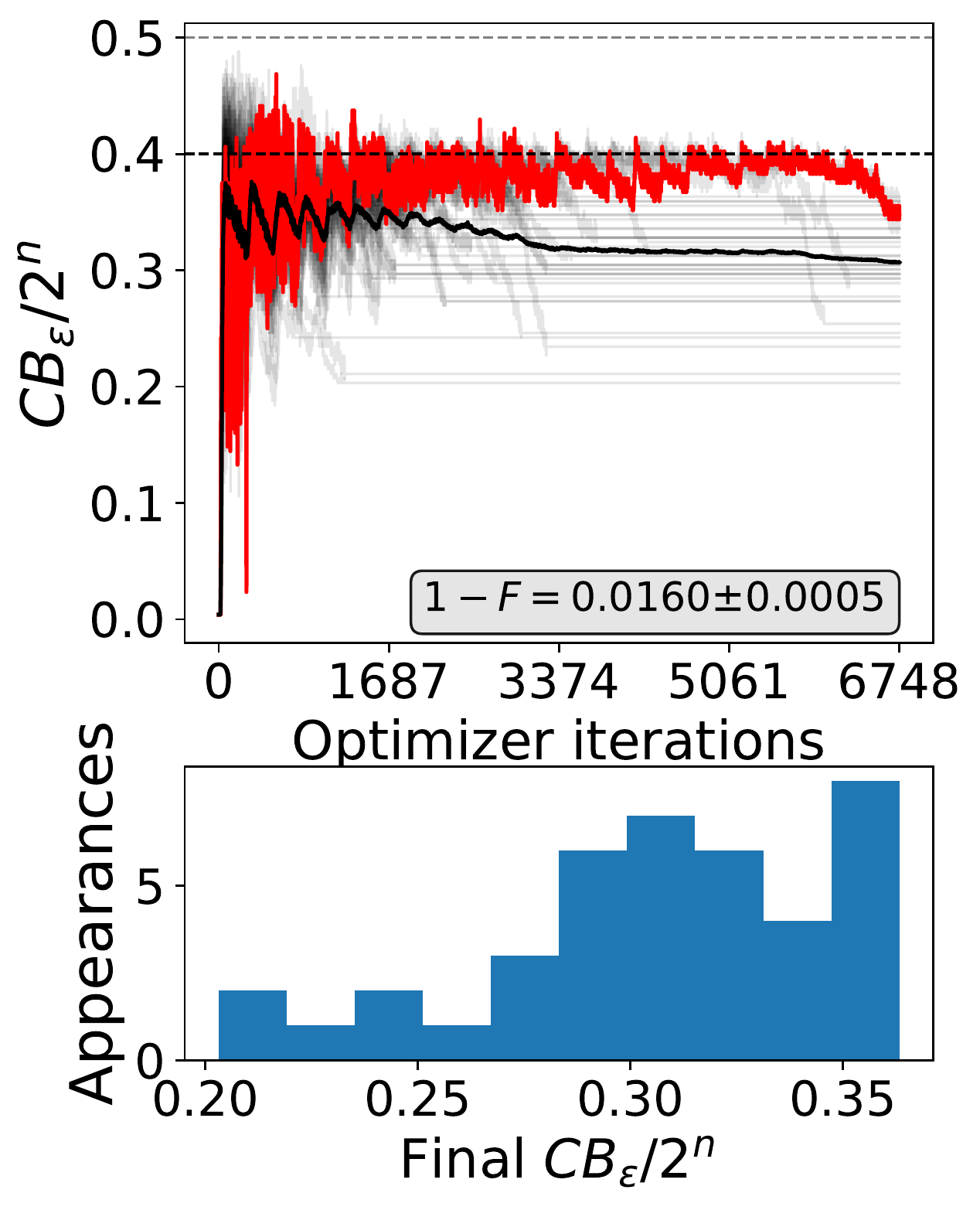}
\label{fig.soft_a}
}
\subfloat[8 qubits, $\varepsilon = 0.03$]{
\includegraphics[width=.45\linewidth]{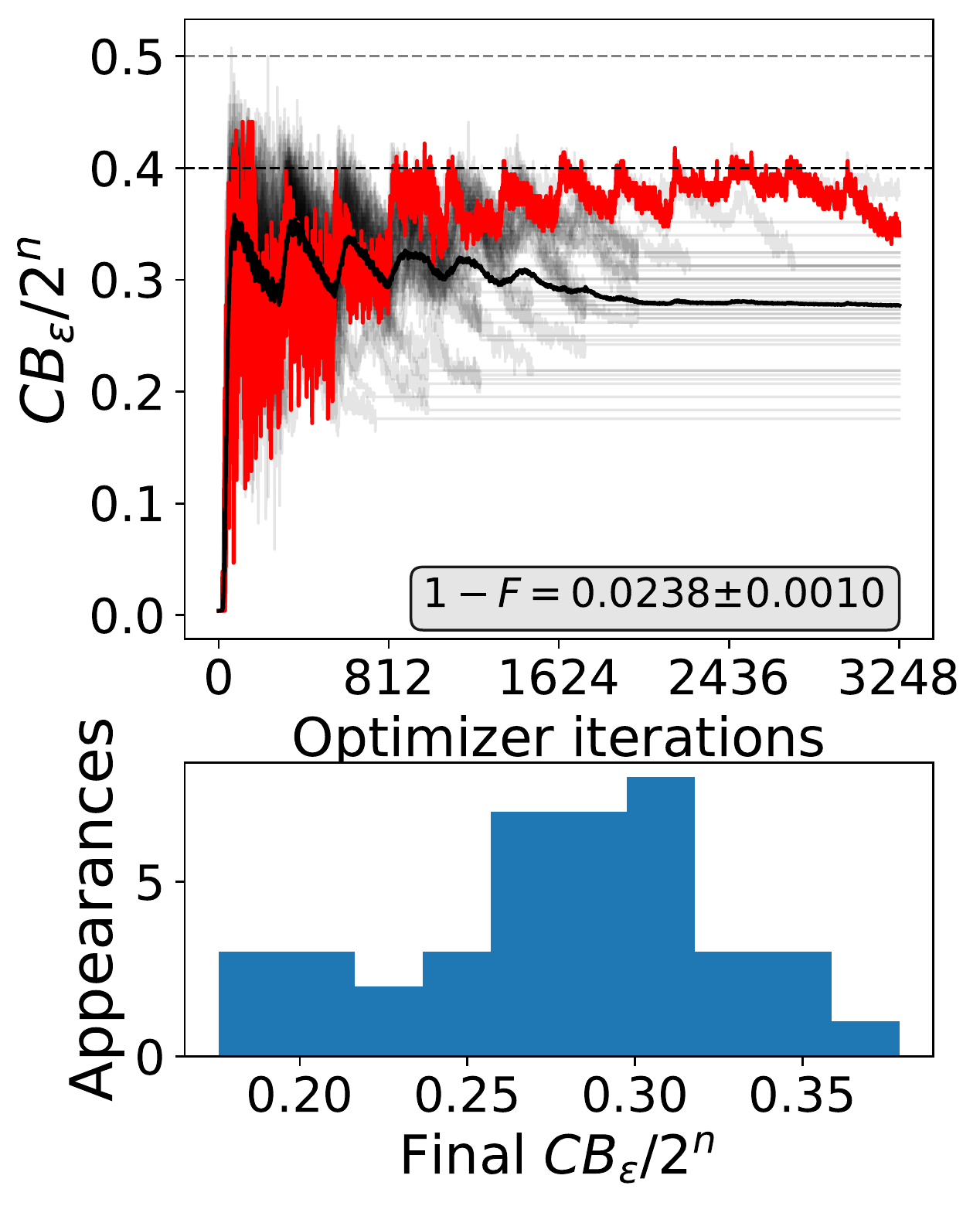}
\label{fig.soft_b}
}

\subfloat[8 qubits, $\varepsilon = 0.05$]{
\includegraphics[width=.45\linewidth]{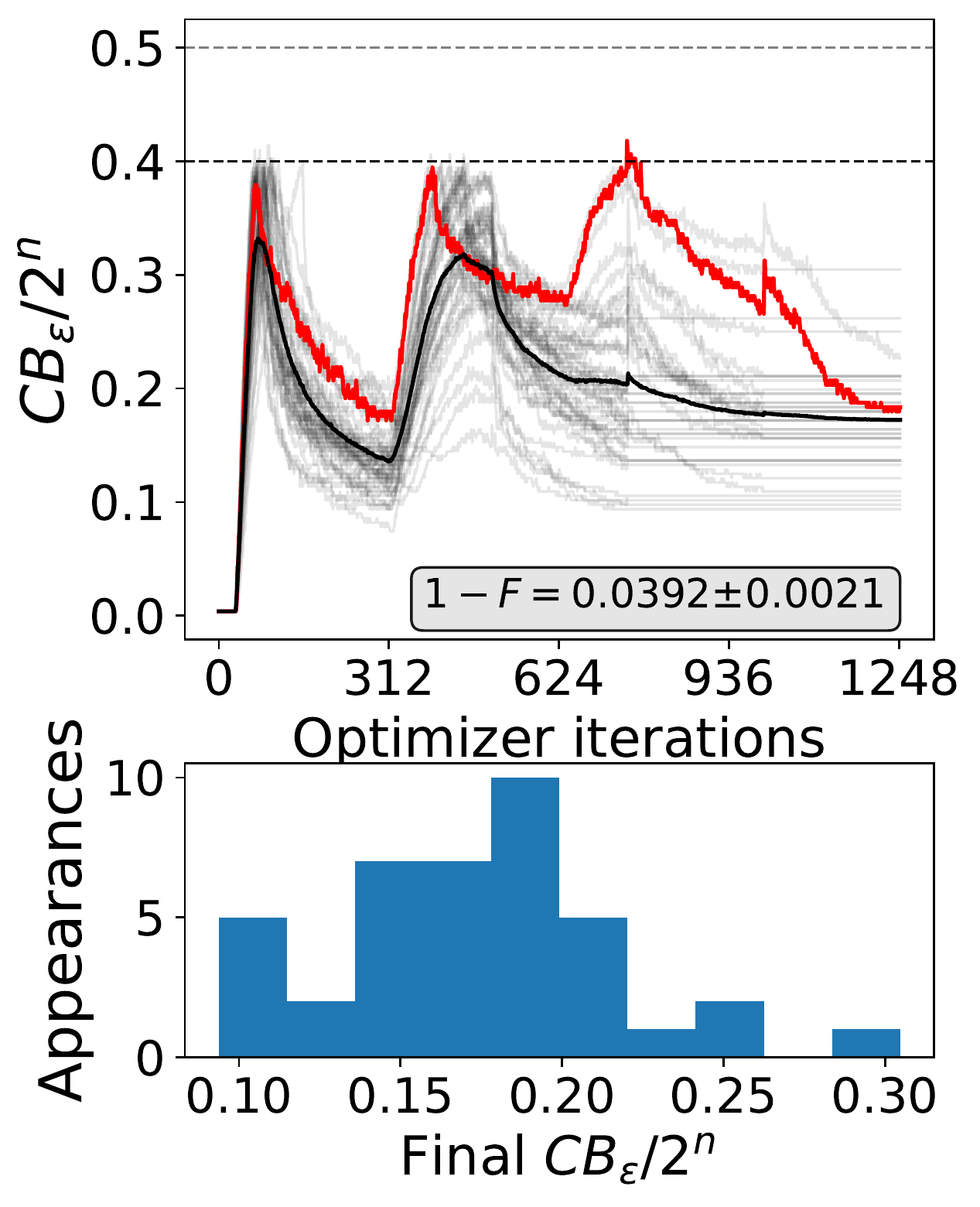}
\label{fig.soft_c}
}
\subfloat[10 qubits, $\varepsilon = 0.03$]{
\includegraphics[width=.45\linewidth]{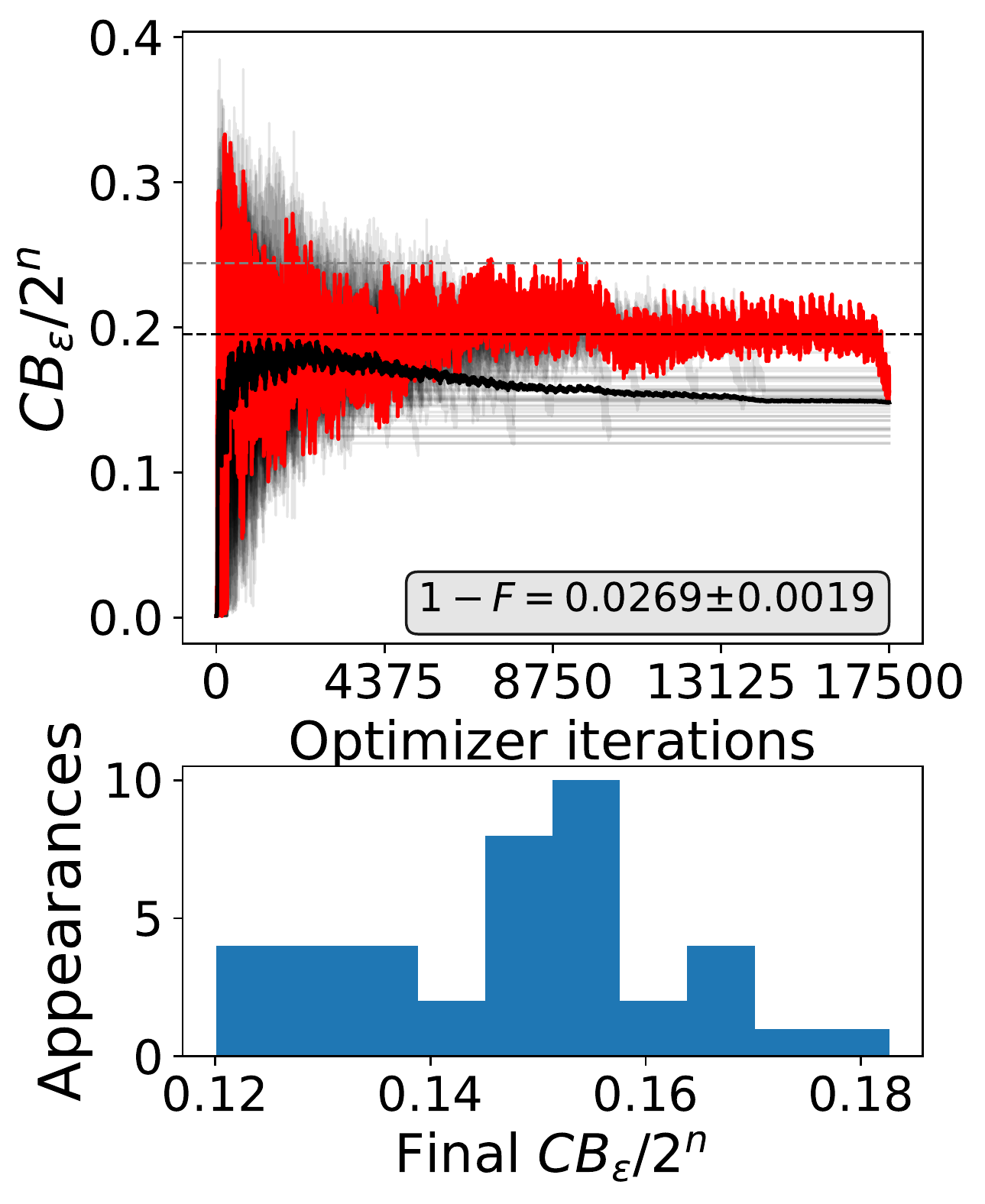}
\label{fig.soft_d}
}

\subfloat[10 qubits, $\varepsilon = 0.08$]{
\includegraphics[width=.45\linewidth]{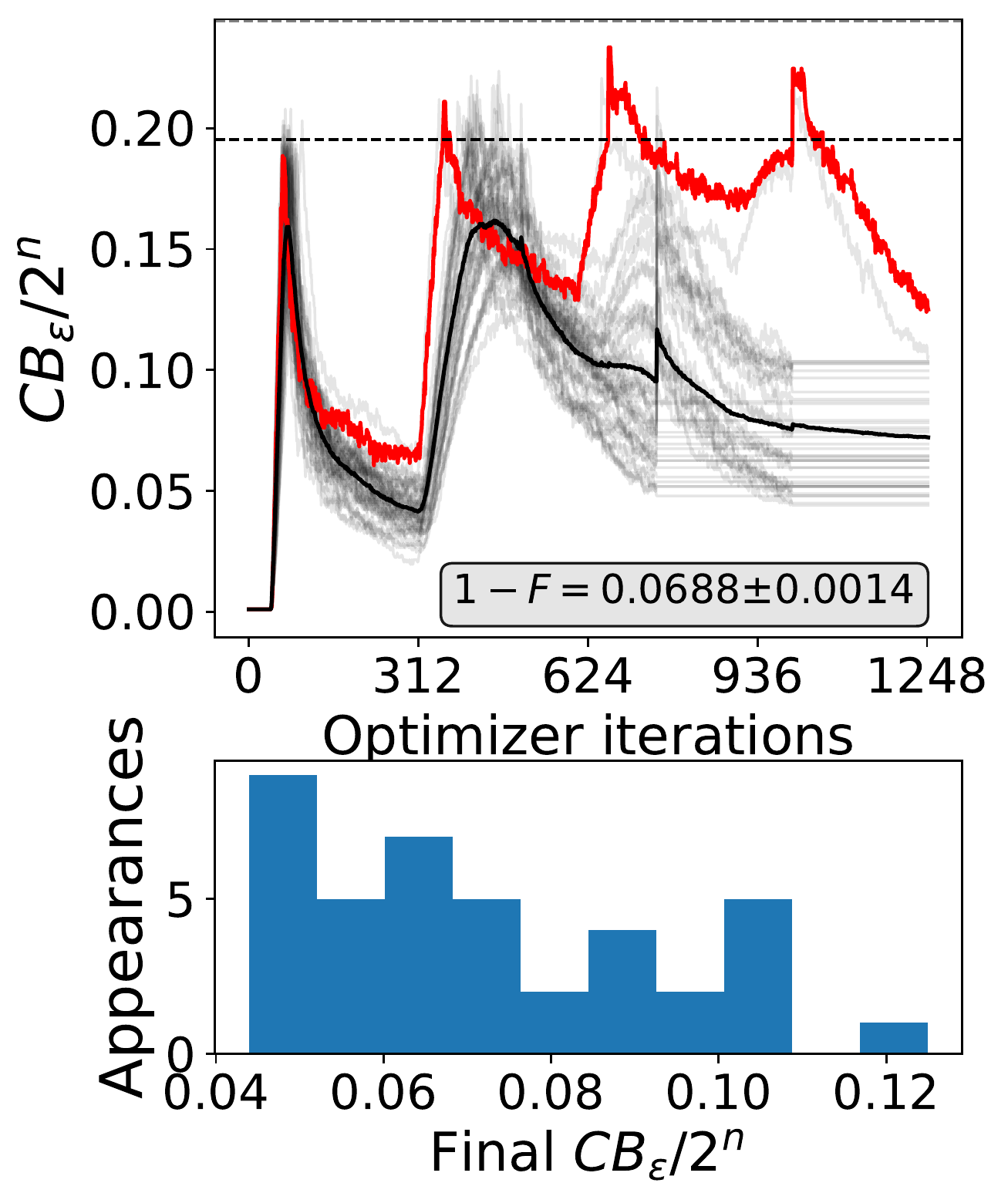}
\label{fig.soft_e}
}
\subfloat[10 qubits, $\varepsilon = 0.13$]{
\includegraphics[width=.45\linewidth]{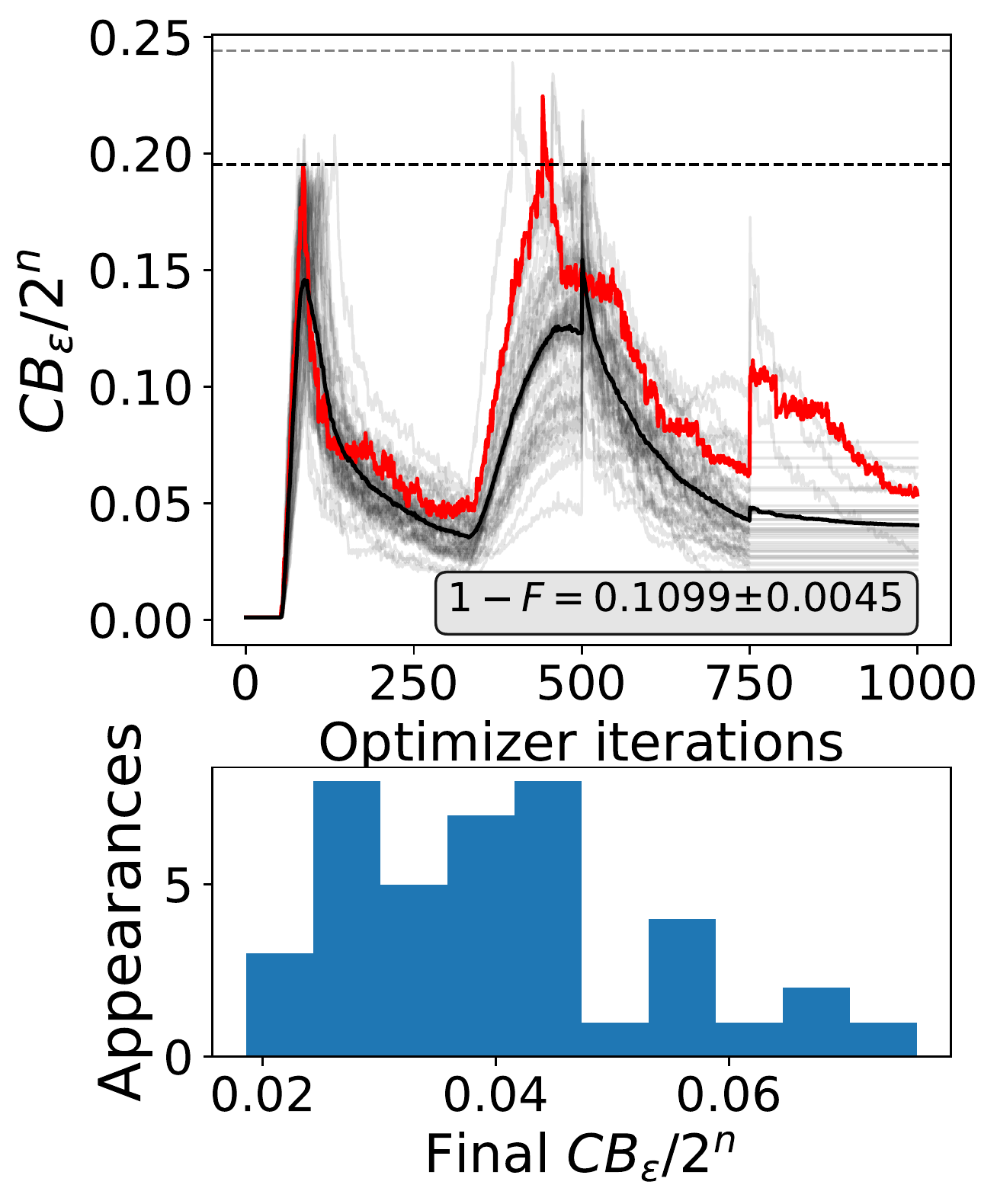}
\label{fig.soft_f}
}

\subfloat{
\includegraphics[width=.7\linewidth]{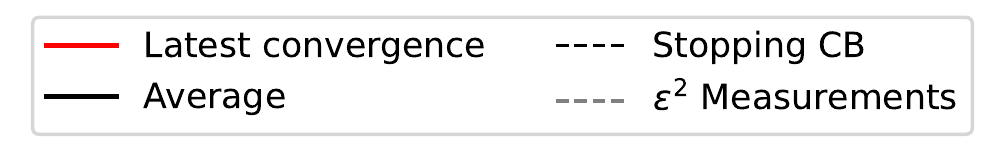}
}
\caption{$\cbe$-ranks as a function of the iteration step in reducer optimization for soft activation. The figures correspond to 40 random instances of the parameters to defining the circuit to be chopped, all of them depicted in black. The thick black lines are the mean values. The red line is the instance with longest optimization. In each graph the maximum $\cbe$-rank allowed during the process and the final fidelity $F$ are detailed. In the bottom row there is an histogram of the final achieved $\cbe$-rank. }\label{fig.soft}
\end{figure}

\begin{figure}
\subfloat[8 qubits, $\varepsilon = 0.02$]{
\includegraphics[width=.45\linewidth]{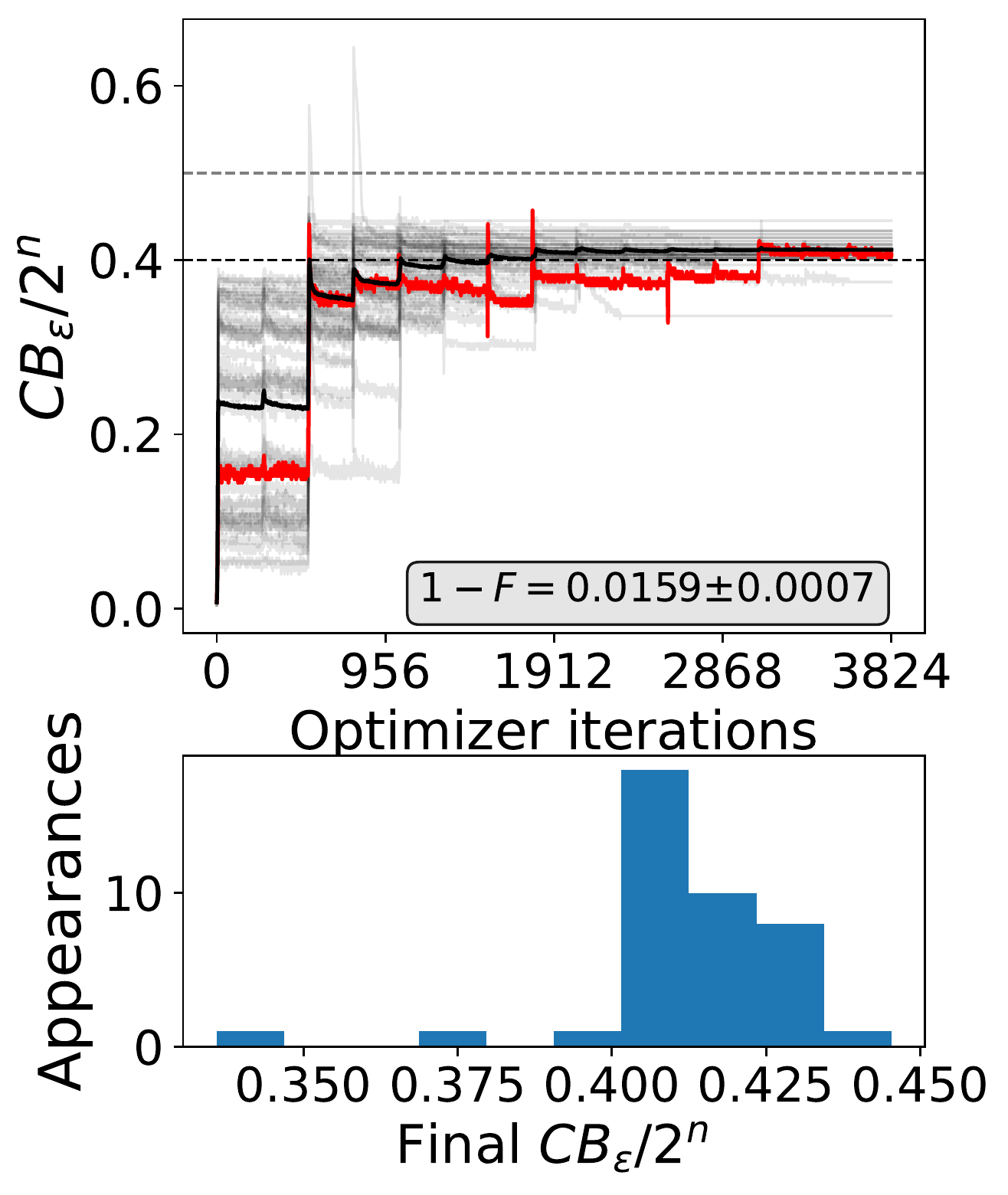}
\label{fig.parametric_a}
}
\subfloat[8 qubits, $\varepsilon = 0.05$]{
\includegraphics[width=.45\linewidth]{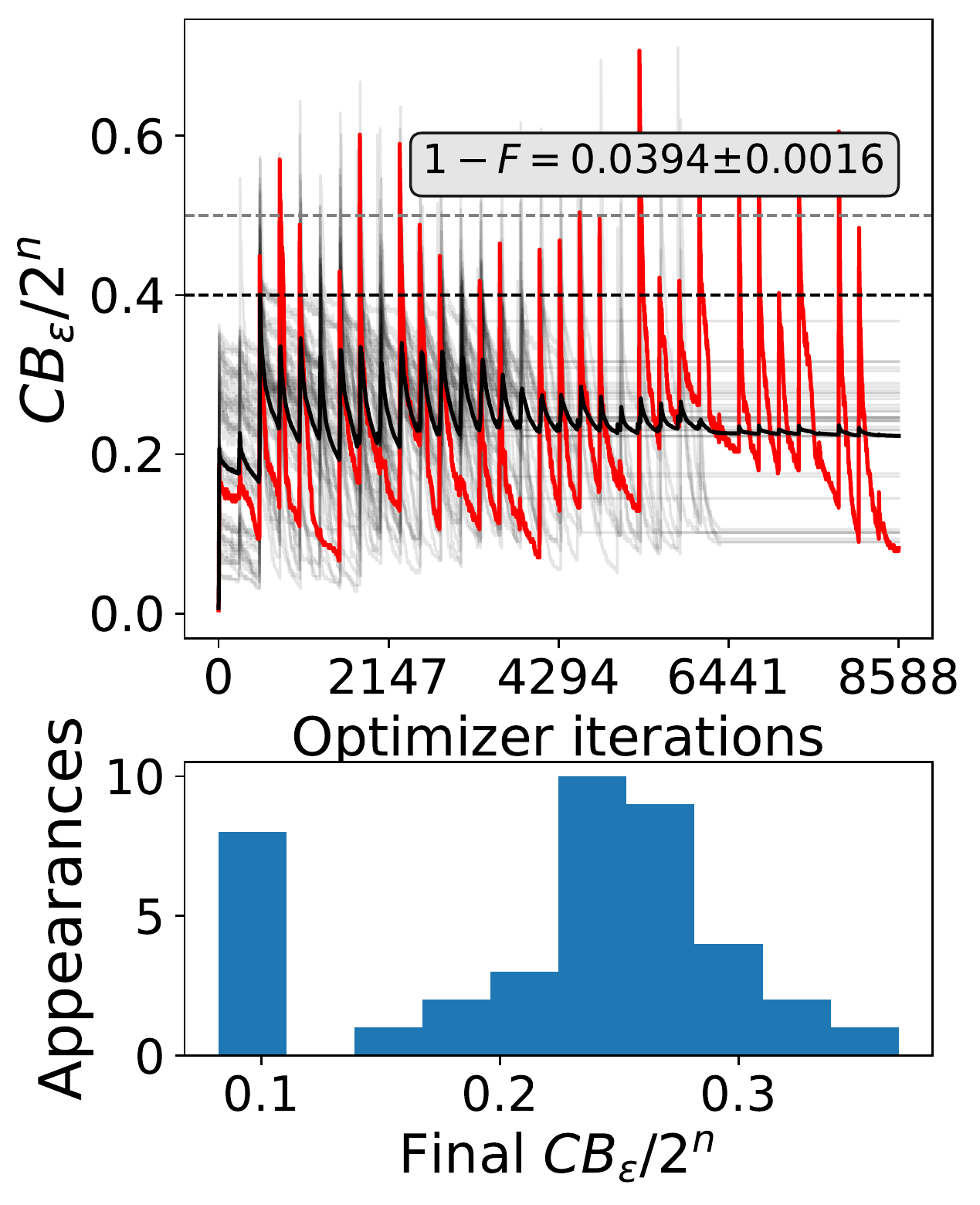}
\label{fig.parametric_b}
}

\subfloat[8 qubits, $\varepsilon = 0.08$]{
\includegraphics[width=.45\linewidth]{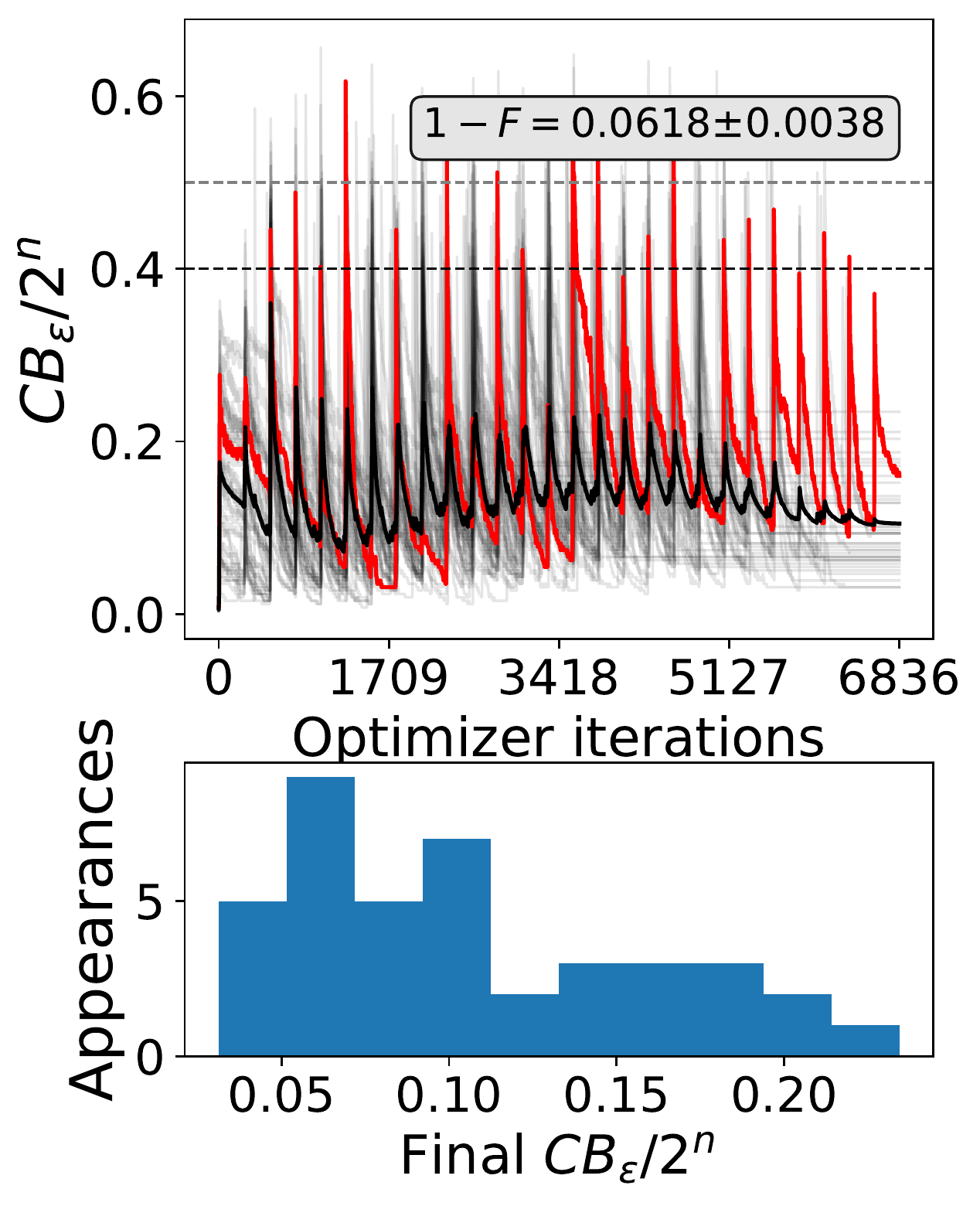}
\label{fig.parametric_c}
}
\subfloat[10 qubits, $\varepsilon = 0.05$]{
\includegraphics[width=.45\linewidth]{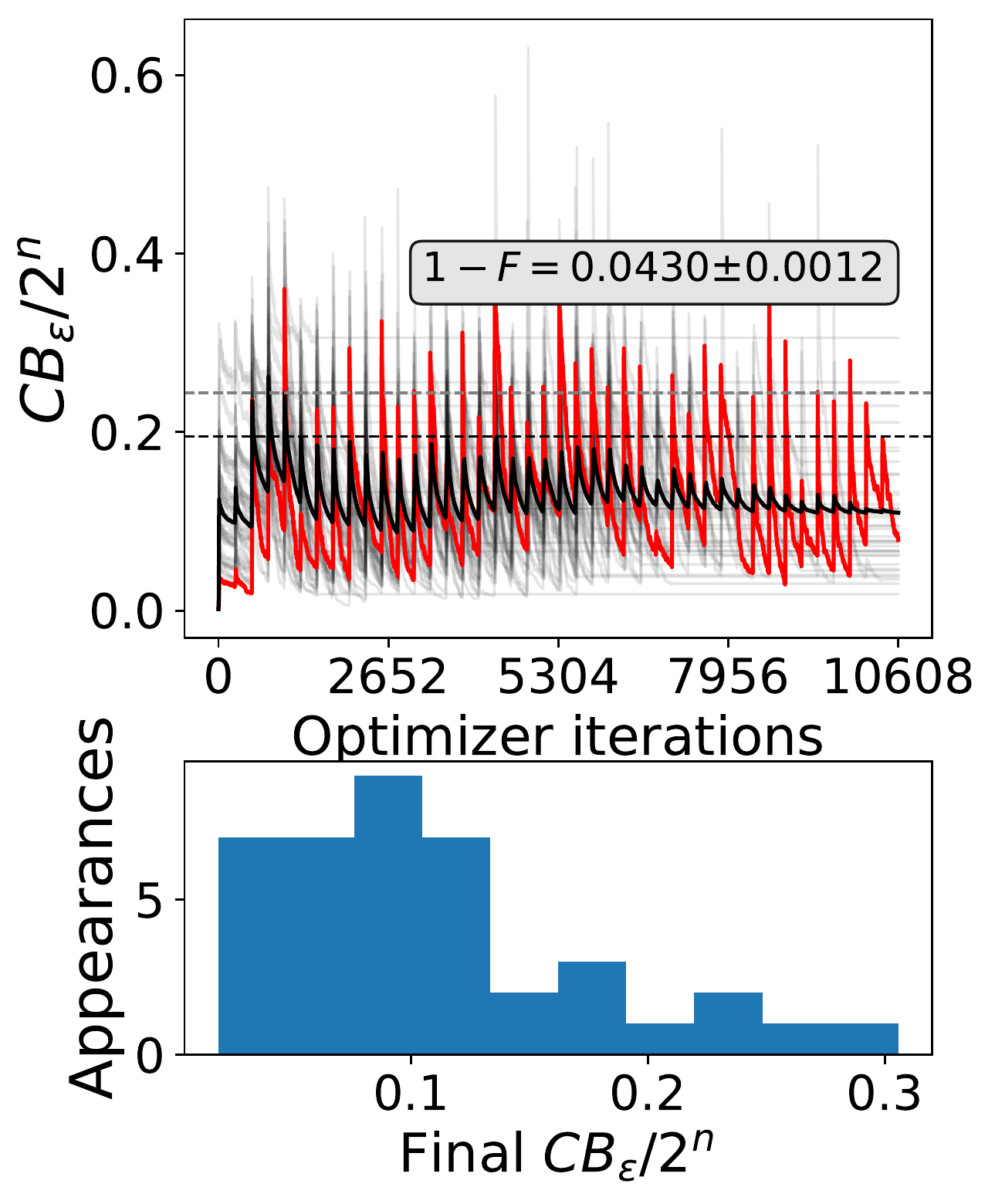}
\label{fig.parametric_d}
}

\subfloat[10 qubits, $\varepsilon = 0.08$]{
\includegraphics[width=.45\linewidth]{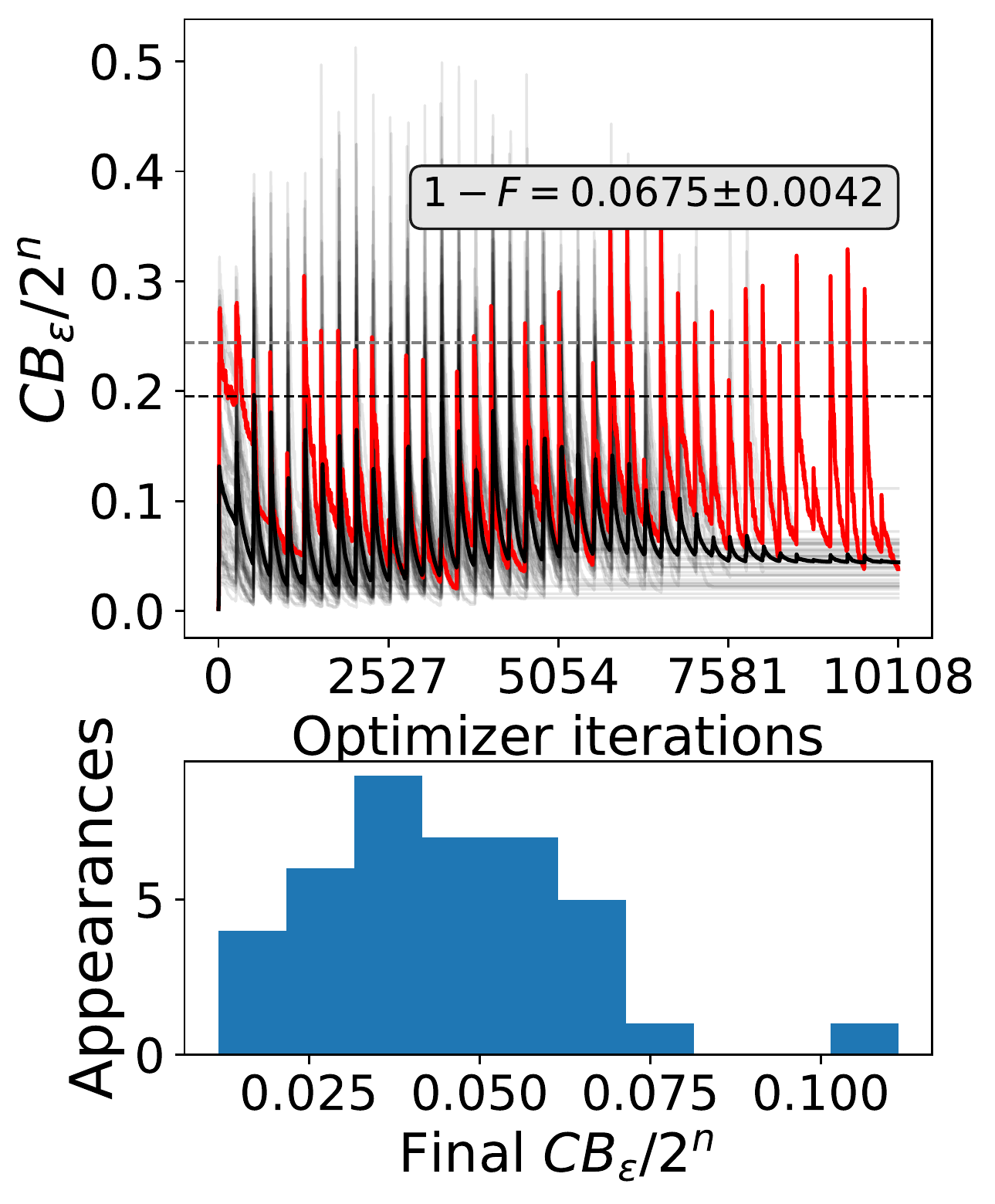}
\label{fig.parametric_e}
}
\subfloat[10 qubits, $\varepsilon = 0.13$]{
\includegraphics[width=.45\linewidth]{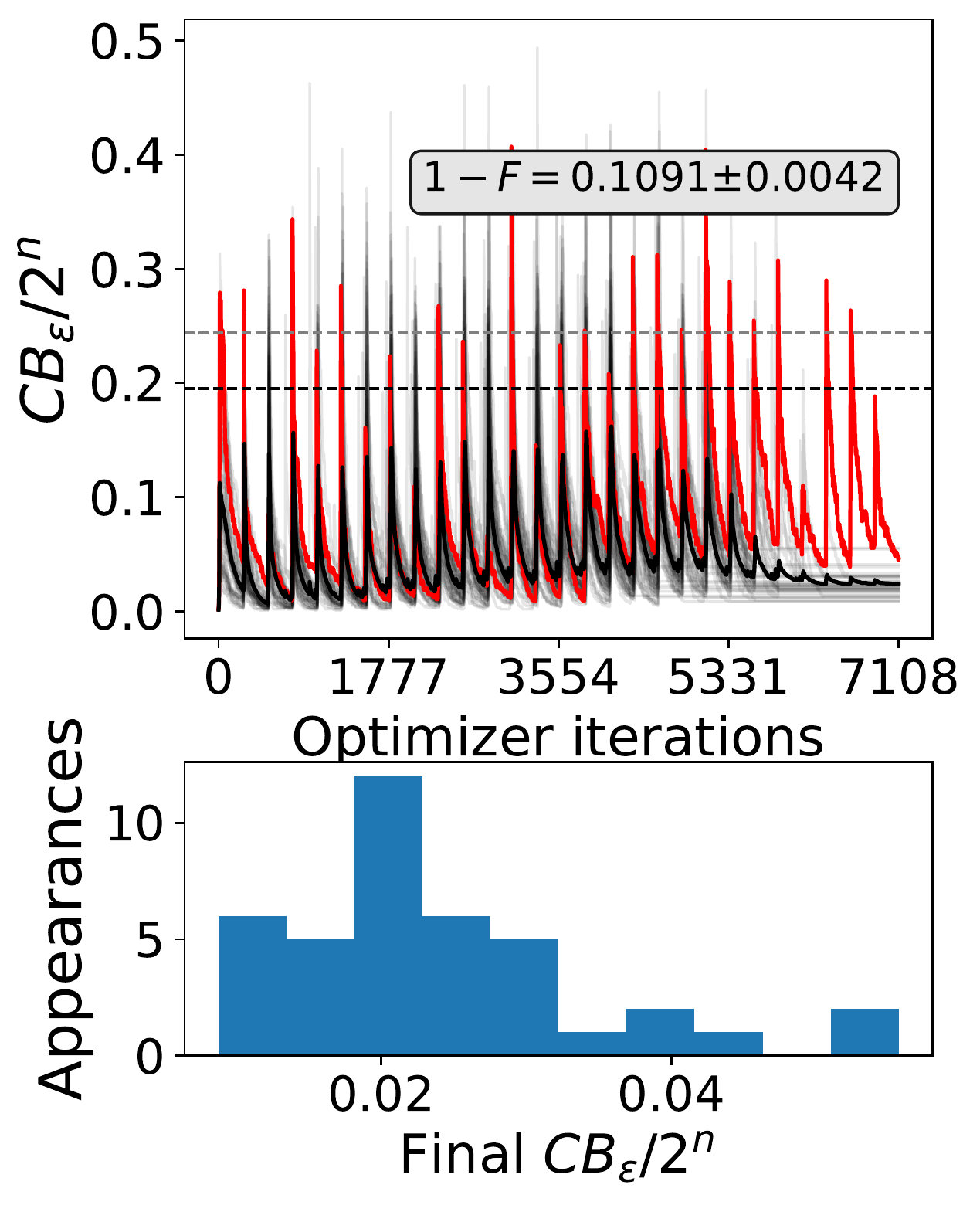}
\label{fig.parametric_f}
}

\subfloat{
\includegraphics[width=.7\linewidth]{figures/tfim_init/soft/legend.pdf}
}
\caption{
$\cbe$-ranks as a function of the iteration step in reducer optimization for parametric activation. The figures correspond to 40 random instances of the parameters to defining the circuit to be chopped, all of them depicted. The thick black lines are the mean values. The red line is the instance with longest optimization. In each graph the maximum $\cbe$-rank allowed during the process and the final fidelity $F$ are detailed. In the bottom row there is an histogram of the final achieved $\cbe$-rank.}
\label{fig.parametric}

\end{figure}

The results of the experiments conducted are summarized in Figures~\ref{fig.soft} and~\ref{fig.parametric}, for several instances of the circuits to be chopped. The results reflect in the top row the values of the $\cbe$-rank through the entire two-stage nested process, including both activation of $U_1$ circuit and optimization of the reducer $R$. All instances are depicted in gray, with the average in bold dark, and the result with latest convergence in red. In the bottom row we depict the final $\cbe$-rank values histogram. In all cases, the average fidelity $F = \vert \braket{\psi}{\psi_K}\vert^2$ between the states created by the circuits with and without chop is computed as a measure to benchmark the quality of the approximation. In these experiments, we lower the depth requirements of a circuit from $40$ to $24$ entangling layers. 

The soft activation is depicted in~\Cref{fig.soft}, for experiments with 8 and 10 qubits, and several values of $\varepsilon = \{0.02, 0.03, 0.05, 0.08, 0.13 \}$. The behavior of the $\cbe$ rank along the optimization process present different properties mainly depending on $\varepsilon$. 

For small $\varepsilon$, mainly~\Cref{fig.soft_a} and~\ref{fig.soft_d}, the $\cbe$-rank is kept approximately constant during the entire optimization process. This can be interpreted as follows. The circuit is activated until $\cbe = CB^{(M)}$. At this point, the optimization of the reducer starts. However, the high required accuracy imposes a constraint that prevents $\cbe$-rank to decrease significantly. After convergence, the circuit continues its activation. The difference to $CB^{(M)}$ is so small the optimizer must come into action immediately again. This process is repeated many times, following a slow activation and optimization path. As the value for $\varepsilon$ increases, the optimizer is more efficient in reducing the $\cbe$-rank of the state at the chop, there is much more room to look for improvements. A transition is observed in~\Cref{fig.soft_b}, where still the slow path observed previously is appreciable with less small steps. In the case of larger $\varepsilon$, as in ~\Cref{fig.soft_c}, \ref{fig.soft_e} \ref{fig.soft_f}, the activation can be conducted through an easier path. After the maximum allowed $\cbe$ rank is obtained, the optimizer can effectively decrease this value. The circuit continues its activation and follows a long path until the maximum value is again exceeded. This leads to an activation process with less interruptions. 

Another element to appreciate in the activation is that estimations of the $\cbe$-rank exceeding $CB^{(M)}$ contribute to reduce the complexity of the full algorithm. The $\cbe$-rank that dominates the complexity is the final value after optimization. Any intermediate value is only required for optimization process.
In addition, even for large $\cbe$-rank, its estimation with some confidence requires only polynomially many samples per evaluation, thus the process is efficient. This phenomenon dominates in~\Cref{fig.soft_d}, where the estimated $\cbe$ ranks during the process exceed the stopping criteria and the number of measurements times $\varepsilon^2$. 
Finally, observe that the final distribution of $\cbe$-ranks have a great dependency on the value of $\varepsilon$. The larger $\varepsilon$ is, the smaller values the final $\cbe$-rank takes, concentrating away from the stopping criteria. 

The parametric activation is summarized in~\Cref{fig.parametric}. In this case the abrupt activation of few parameters implies a large change in the $\cbe$-rank. 
Periodic optimizations without the requirement to reach $CB^M$ are conducted to ease the task. The first iterations of the optimizer may lead to configurations where the $\cbe$ rank take high values, see in particular~\Cref{fig.parametric_b}, \ref{fig.parametric_c}, \ref{fig.parametric_d}, \ref{fig.parametric_e}, \ref{fig.parametric_f}. These higher values may compromise the activation process if they are too large to estimate the $\cbe$ rank with the assigned measurement budget. It is not guaranteed that this heuristic method succeeds in these instances, but results show a solution.  The performance depends on the particular instance to solve. 

The special case of small $\varepsilon$ for 8 qubits is depicted in~\Cref{fig.parametric_a}. The room the optimizer has to reduce the $\cbe$ rank is small, so the process tends to concentrate around the stopping criteria. In this case, the optimization is not successful for all instances, which are the ones ending above the stopping criteria.

The choice of the loss threshold $\varepsilon$ has important consequences for optimization. The soft activation is more sensitive to $\varepsilon$, and its value affects both the difficulty of the optimization and the final $\cbe$-rank. The smaller $\varepsilon$ is, the more iterations are required for optimization. In fact, unstabilities are more strongly observed when $\varepsilon$ is small, and thus the optimization procedure is more costly. In case $\varepsilon$ is too small, there is a chance we cannot find a set of parameters that reduces the $\cbe$-rank. This means that the exploration of the landscape reaches the area where $\cbe$-ranks cannot be estimated. 

Another element to take into account is the fidelity between the state of the circuit after the chop, and the target state generated by $R U_1$. This quantity is added to each graph. The final values of the fidelity grow with $\varepsilon$, not overcoming the result in~\Cref{le.error}. Thus, there is a direct relationship between two quantities, and it is possible to choose $\varepsilon$ prior to the optimization process by taking as a reference the allowed disagreement in the final output distributions. Smaller $\varepsilon$ values also trigger more costly optimization processes and potentially larger values of $\cbe$-rank. These two aspects define the final computational cost of our algorithm. Overall, there is a trade-off between achievable precision and computational cost.

The numerical experiments here presented aimed at exploring two different possible activation strategies. These first results may shed some light on future explorations of an optimal activation recipe. A possible direction to explore a better activation process is to create a smooth path that follows the one described by the soft activation, but that is directly implementable on the parameters of a quantum computer, for example using adiabatic approaches. 

\section{Extensions of reduce\&chop method}\label{sec.extension}

\subsection{Decision problems, sampling, expectation values}

\paragraph{Decision problems:}The algorithm we presented allows us to compute the probability of measuring a given bitstring.
This capacity suffices to address circuits which decide BQP decision problems such as deciding if $\vert\bra{0} U \ket 0\vert^2$ is above 2/3 or below 1/3 (under the promise it is one of these cases). The attractive feature of BQP decision problems is that the {\tt Yes} and {\tt No} instances measurement probabilities are separated by a large gap. Then, we can set $\varepsilon$ in our $\cbe$ robustness to a large value, which makes the procedure much more likely to succeed.

\paragraph{Sampling problems:}Another important problem is sampling from the state $U|0\rangle$.
The reduce\&chop does not allow a simple method for this 
but a heuristic is possible. The problem of sampling from $P(x)$ can be addressed using a Metropolis-like algorithm in a similar manner to the one considered in Ref.~\cite{bravyi2019simulation}. Only the values $P(x)$ for given $x$ are needed. A big shortcoming is starting in a bitstring $x$ with small $P(x)$, leading to exponentially long mixing times. To mitigate this in some cases we can use the assumption of low-$\cbe$-rank middle states. Then, there are only few possible final states $\{ U_2R^\dagger |j\rangle\} $, possibly with high $\cbe$ rank. If the support is localized, then a simple sampling is enough to produce a good approximation of the output state. Otherwise, prioritizing samples from that set of states is likely to improve the convergence of the Metropolis-like algorithm.

\paragraph{Estimating expectation values:} 
Estimating the expectation value of a state with respect to a given observable is easy only in certain cases, i. e. if the observable is related to a polynomial sum of computational-basis projection, or any other basis with low-depth basis conversion from the computational basis.
For other observables, e.g. related sums of Pauli strings, the methods are not efficient, and one is probably best relying on the heuristic sampling method. If sampling from the state is possible, then all diagonal observables are easy, as are all Pauli strings (via local basis change), and sums of Pauli strings (due to linearity).

\subsection{Generalizations}

At present, the entire method relies on recovering one succinct description of the state at the half-way cut: sparse representation in the computational basis. 
One can consider more general schemes where we e.g. utilize overcomplete basis sets (more precisely, unions of basis sets) related to, e.g., stabilizer basis and also classical shadows~\cite{aaronson2004improved, huang2020predicting}.

Further, as mentioned, multiple cuts are possible as well, see~\Cref{app:feynman}; in this case one faces two challenges:\begin{inparaenum}[(1)] the exponential growth of terms to consider in the number of cuts, and \item the fact that the middle terms will include the evaluation of ``cross-terms''  $\langle i| U_2 | j \rangle$ in which case the estimation of total error becomes more involved. \end{inparaenum} We leave such consideration for future work.

\subsection{Beyond simulation of deeper circuits}\label{sec.beyond}

Finally, one can drop the initial objective of simulating a given fixed circuit, but rather using shallower circuits directly to solve a desired problem.
Consider for the moment the standard VQA problem to find $ \textup{argmin}_{\bm\phi} \bra 0 U^\dagger(\bm\phi) H U(\bm\phi) \ket 0$. In our approach we modify this to a two stage optimization
\begin{equation}
\underset{\bm\theta, \bm\phi}{\rm argmin} \Vert H U_2(\bm\phi) R^\dagger(\bm\theta) R(\bm\theta) U_1(\bm\phi) \ket 0 \Vert ,
\end{equation}
subject to the additional constraint that $ R^\dagger(\bm\theta) |  U_1(\bm\phi) \ket 0 $ is a low-$\cbe$ state. We assume for the ease of presentation and without loss of generality that $H$ is positive semidefinite. 
At this point, we may as well merge the parameters to one parameter family, with the collective optimization problem:
\begin{equation}
 \underset{\substack{ \bm\phi \atop   \text{s.t.} \cbe(  W_1(\bm\phi) \ket 0) < CB^{(M)}} }{\rm argmin} \Vert  H W_2(\bm\phi)   W_1(\bm\phi) \ket 0\Vert
\end{equation}
where we merged the reducer and Ansatz into one unitary, and the optimization now runs over the complete parameter space. The constraint of low $\cbe$-rank at the cut is the connection to reduce\&chop.
Notice that adding constraints in an optimization problem can ease the procedure and find better solutions, although the global optima is at best equal. In this case, however, we expect the constraint to make the optimization problem more challenging, an issue added to others already known~\cite{mcclean2018barren}. This optimization scheme also allows in principle to obtain quantum states out of reach for (shallow) unitary quantum evolutions, due to the intermediate measurement.

This optimization scheme also offers protection against noise, as compared to standard unitary ansatzes. For unitary ansatzes, noise occurs during the entire circuit and decreases purity of the quantum state stored in the processor. For reduce\&chop optimization, the state is pure immediately after the cut, and decoheres only due to the noise in the second half of the circuit. Such decoherence for half a circuit is smaller as compared to the entire circuit. Optimization is also more robust to errors than the simulation of deep circuits discussed in~\Cref{sec.cb_rank}.

\section{Conclusions}\label{sec.conclusions}

This work explores the possibilities of using shallower circuits to perform quantum computations specified by deeper circuits. The approach is in spirit related to other ``circuit-cutting'' methods~\cite{marshall2022high, huembeli2022entanglement, peng2020simulating}, mostly concerned with reducing the qubit numbers and not the computational depth. The method we built has in general exponential overheads, but we develop new heuristic machinery to circumvent identified bottlenecks. 

Since the method is mostly heuristic, we performed small-scale numerical experiments to asses the feasibility of the method. As an illustrative example, we have shown how our method allows us to reduce the depth requirements for a physically motivated Ansatz from 40 to 24 layers. The results are promising, however the scaling of the approach is hard to assess. It is unclear if real-world interesting quantum computations are immediately accessible with reduce\&chop. 

There exist several directions to further investigate the applicabiliity of reduce\&chop. It can serve as a subroutine for more sophisticated problems requiring deeper circuits. As a first generalization, the ideas in this work can be taken as an inspiration for succinct descriptions of quantum states, in the spirit of stabilizer states. Versions of reduce\&chop for variational algorithms are to be explored as well. 

We envision this work to be a first step towards the exploration of tools to lower the depth requirements of a quantum circuit to be executed in a given hardware. This aspect is understudied, despite its applicability in near-term devices with limited available depth, in contradistinction to other hardware limitations already explored.

\acknowledgements
APS would like to thank Leonardo Lenoci for his help to conduct the numerical simulations in this paper. We also acknowledge Stacey Jeffery for useful discussions. This work has received support from the European Union’s Horizon Europe program through the ERC StG FINE-TEA-SQUAD (Grant No. 101040729). The authors also acknowledge support from the Quantum Delta NL program. This work was in part supported by the Dutch Research Council (NWO/OCW), as part of the Quantum Software Consortium program (project number 024.003.037). This work is also funded by the European Union under Grant Agreement 101080142 and the project EQUALITY. This publication is part of the ‘Quantum Inspire – the Dutch Quantum Computer in the Cloud’ project (with project number
[NWA.1292.19.194]) of the NWA research program ‘Research on Routes by Consortia (ORC)’, which is funded by
the Netherlands Organization for Scientific Research (NWO). 

\bibliography{bibliography}

\onecolumngrid

\appendix

\section{Reduce\&chop algorithm for many subcircuits}\label{app:feynman}
\begin{figure}[h!]
    \centering
\includegraphics[width=\linewidth]{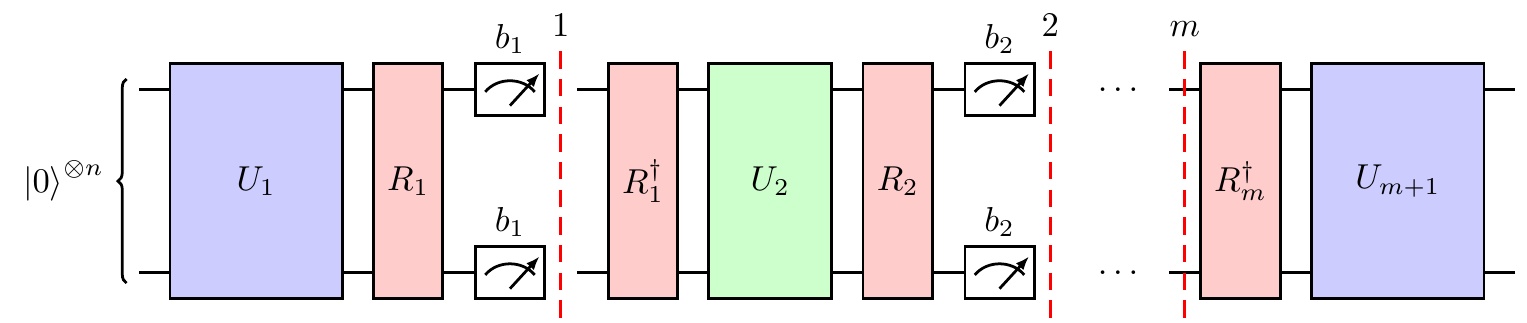}
\caption{Scheme for the sequential splitting of a quantum circuit including variational reducers $R$. Red dashed lines stand for the points where the measurements are applied. The $U_i$ pieces are determined by the algorithm to run, while the $R_i$ ones are reducers that aim to minimize the $\cbe$-rank at the chop. The reducers are applied before the chop and reversed after. The total depth between red lines cannot exceed the hardware limits.}
    \label{fig.cuts}
\end{figure}

We consider a circuit $U$ with depth $D$, to be evaluated using a quantum computer restricted to computations up to depth $d$. The strategy to simulate the execution of a $D$-depth computation is as follows. The circuit $U$ is sequentially split in $m$ chops as
\begin{equation}
    U = U_{m+1}U_{m}\dots U_1,
\end{equation}
in such a way that the depth of each $U_i$ is strictly smaller than $d$. In fact, we need it to be considerably smaller to make room for the reducer $R$. In the spirit of Feynman's sum-over-path approach, we can rewrite the probability $P(x)$ as
\begin{equation}\label{eq:prob_def}
    P(x) = \vert\bra{x}U_{m + 1} U_m\dots U_1\ket{0}\vert^2 = \left\vert\sum_{b_1, b_2, \dots, b_{m}} \bra{x}U_{m+1}\ket{b_{m}} \bra{b_m}U_{m}\ket{b_{m - 1}} \ldots \bra{b_1}U_{1}\ket{0}\right\vert^2,
\end{equation}
where the sum is obtained by inserting the resolutions of the identity indexed by $b_i$. The reducers are introduced in each chop as
    \begin{equation}
\small
    P(x) =
    |\bra{x}U_{m+1}R^\dagger_mR_mU_{m}\ldots R_1U_1\ket{0}|^2 = \left|\sum_{b_1, b_2, \dots, b_{m}} \bra{x}U_{m + 1}R^\dagger_{m}\ket{b_m}
    \bra{b_m}R_{m}U_{m}R^\dagger_{m - 1}\ket{b_{m - 1}} \ldots 
    \bra{b_1}R_{1}U_{1}\ket{0}\right|^2,
\end{equation}
See~\Cref{fig.cuts} for a graphical scheme of the algorithm here depicted. Notice that the middle chops include two different reducers, thus the available depth in the corresponding stage is smaller. This is solved by selecting a division setting with depths of $U_i$ smaller in the intermediate steps.

\section{Proof of \Cref{approx_cb_rank_state}} \label{app:proof}

Let $\mathcal{S}^{(K)}$ be the set of states with at most $K$ non-zero coefficients. Let us take some state $\ket{\phi} = \sum_{i = 0}^{N - 1}\beta_i\ket{i} \in \mathcal{S}^{(K)}$ .If $\pi$ and $\sigma$ are permutations such that $|\alpha_{\pi(i)}| \geq |\alpha_{\pi(i + 1)}|$ and $|\beta_{\sigma(i)}| \geq |\beta_{\sigma(i + 1)}|$ for $i \in \{0, \dots, N - 2\}$, then
\begin{flalign}
|\braket{\psi}{\phi}| &= \left|\sum_{i = 0}^{N - 1}\alpha^*_i\beta_i\right| \leq \sum_{i = 0}^{N - 1}|\alpha^*_i\beta_i| = \sum_{i = 0}^{N - 1}|\alpha_i||\beta_i|\\\nonumber
&\leq \sum_{i = 0}^{N - 1}|\alpha_{\pi(i)}||\beta_{\sigma(i)}| = \sum_{i = 0}^{K - 1}|\alpha_{\pi(i)}||\beta_{\sigma(i)}|\\\nonumber
&\leq \sqrt{\sum_{i = 0}^{K - 1}|\alpha_{\pi(i)}|^2\sum_{i = 0}^{K - 1}|\beta_{\sigma(i)}|^2} = \sqrt{\sum_{i = 0}^{K - 1}|\alpha_{\pi(i)}|^2},
\end{flalign}
where we applied triangle inequality, rearrangement inequality, and Cauchy-Schwartz inequality respectively. Hence,
\begin{equation}
\max_{\ket\phi \in \mathcal{S}^{(K)}}\left(\vert\braket{\psi}{\phi}\vert\right) = \sqrt{1 - \sum_{i = 0}^{K - 1}|\alpha_{\pi(i)}|^2}
\end{equation}
is achieved for the state
\begin{equation}
\ket{\phi} =  \left(\sum_{j = 0}^{K - 1}|\alpha_{\pi(j)}|^2\right)^{-1/2}\; \sum_{i = 0}^{K - 1}\alpha_{\pi(i)}\ket{\pi(i)}.
\end{equation}
\hfill $\square$

\section{Proof of~\Cref{le.CB^{(M)}easure}}\label{ap:cb_rank}
The proof consists in transforming the estimation of the $\cbe$-rank of a state into a binary process to use bounds in the binomial distribution. 

First, $2M$ samples are collected and divided into two different sets of size $M$ at random. Since measurements are random, a valid split is the first and second halves. This returns two collections of bitstrings and frequencies. This returns the set of triplets
\begin{equation}
R = \{(x_i, n_i, m_i)\}, 
\end{equation}
where $x_i$ is a bitstring, $m_i$ is the frequencies of $x_i$ in the first sampling and $n_i$ is the frequency of $x_i$ in the second sampling. For convenience, the set is ordered in descending order according to $m_i$. Notice that if the number of samples is sufficiently large $m_i \sim n_i$. By definition, $\sum_i m_i =M$, but $\sum_i n_i \neq M$. We define now a set $S(k) \in R $  as 

\begin{equation}
S(k) = \{x_i \; \vert i \leq k \}, 
\end{equation}

The binary process is then defined with respect to the set $S(k)$. A sample from the second set with output $y$ is assigned with the labels
\begin{equation}
B(y) = \left\{ \begin{matrix}
0 & \qquad & y \in S(k) \\
1 & \qquad & y \not\in S(k) 
\end{matrix}\right.
\end{equation}

We select now a 
\begin{equation}
K = N\; \left\vert \; \sum_{i = 1}^{N} m_i > (1 - \varepsilon) M \right. ,
\end{equation}
and compute the number of positive outcomes in the binary process

\begin{equation}
m = M - \sum_{i = 0}^K m_i.
\end{equation}

We now define an hypothesis test to check the probability of obtaining a given estimation of the mean of a binomial process with respect to a given probability. In this case, the probability to test is $\varepsilon$, and the estimate of the mean is $m / M$. First, a necessary condition to be matched is $m > \varepsilon M$. Then, direct application of Hoeffding's inequality~\cite{hoeffding1963probability} allows us to bound this quantity as 
\begin{equation}
{\rm Prob}(m < \varepsilon M) \leq \exp\left( -2 M \left(\varepsilon- \frac{m}{M}\right)^2\right).
\end{equation}

A positive hypothesis test of tells then that the $K$ chosen outcomes are enough to represent the quantum state with accuracy at least $1 - \varepsilon$, with a $\cbe$-rank not exceeding $K$, where the relevant bitstrings are the ones in $S(K)$, yielding the desired result
\begin{equation}
{\rm Prob}(\cbe > K) \leq \exp\left( -2 M \left(\varepsilon- \frac{m}{M}\right)^2\right).
\end{equation}
For the purpose of the proof, this bound is enough, but can be made sharper including Chernoff bound~\cite{chernoff1952measure} and the Kullback-Leibler divergence~\cite{kullback1951information}.~\hfill $\square$

Notice that this method measures the concentration of measurements in few outcomes. The method gives a proper estimation if the actual $\cbe$ rank is small and a large portion of measurements $M$ concentrate in a few possible outcomes. In this case, the frequencies $n_i, m_i$ will be similar since there are enough measurements to resolve them. Many samples concentrate in few outcomes, and only few of them appear in bitstrings with low frequency. The number $m$ is be reasonably small and the bounds apply. On the contrary, if the actual $\cbe$-rank is large, both $m_i$ and $n_i$ are small. In particular, $m_i = 0$ and $n_i = 1$ in most cases. There is no concentration in few outcomes, and the number $m$ is so large that either the bounds do not apply or the failure probability is large. 

\section{Error estimation with respect to number of measurements}\label{app:estimation_error}

The quantum state to be approximated can be represented as
\begin{equation}
\ket\psi = \sum_{i = 0}^{K - 1} \alpha_{\pi(i)} \ket{\pi(i)} + \sum_{i = K}^{2^n - 1} \beta_{\pi(i)} \ket{\pi(i)} = \sqrt{1 - \delta}\ket{K} + \sqrt{\delta}\ket{\bcancel{K}}, 
\end{equation}
where $\pi(\cdot)$ is a permutation ordering the coefficients in descending order according to its absolute value. The state $\ket K$ is a short notation for all the coefficients considered in the $K$ most relevant ones, the estimated $\cbe$ rank, properly normalized. The state $\ket{\bcancel K}$ comprises the rest of them. By definition $\delta < \varepsilon$. The state after chop is represented by
\begin{equation}
\ket{\psi_K} = \sum_{i = 0}^{K - 1} \hat\alpha_{\pi(i)} \ket{\pi(i)} = \sqrt{1 - x}\ket{K} + \sqrt{x} \ket{K^\perp}, 
\end{equation}
where $\hat\alpha_{\pi(i)}$ are estimations of the actual coefficients $\alpha_{\pi(i)}$. The state $\ket{K^\perp}$ has the same support as $\ket K$ in terms of states in the computational basis, but it is orthogonal to it, $\braket{K}{K^\perp} = 0$. Notice that the state here is artificially renormalized with respect to the terms found in the measurement, since all terms with no support in the $K$ most relevant cases have been discarded.

As a preliminary result, we relate the relative fidelity of the states to the Euclidean distance. First, the Euclidean distance is
\begin{equation}
    \Vert \ket\psi - \ket{\psi_K}\Vert^2 = \left(\sqrt{1 - x} -  \sqrt{1 - \delta}\right)^2 + x + \delta = 2\left(1 - \sqrt{(1 - x)(2 - \delta)}\right).
\end{equation}
On the other hand, the relative fidelity is
\begin{equation}
    \vert \braket{\psi}{\psi_K}\vert = \sqrt{(1 - x)(2 - \delta)},
\end{equation}
and rearranging
\begin{equation}
    \vert \braket{\psi}{\psi_K}\vert^2 = \left( 1 - \frac{1}{2}\Vert\ket{\psi} - \ket{\psi_K} \Vert^2\right)^2
\end{equation}

The goal now is to bound the norm of the difference of the states as 
\begin{align}
\Vert \ket\psi - \ket{\psi_K}\Vert^2 & = \sum_{i = 0}^{K - 1} \vert\hat\alpha_{\pi(i)} - \alpha_{\pi(i)}\vert^2 + \sum_{i = K}^{2^n - 1} \vert\beta_{\pi(i)}\vert^2 \leq \\
&\leq \left\Vert \sum_{i = 0}^{K - 1} (\hat\alpha_{\pi(i)} - \alpha_{\pi(i)}) \right\Vert^2 + \varepsilon \leq \\
& \leq \sum_{i = 0}^{K - 1}\left\Vert \hat\alpha_{\pi(i)} - \alpha_{\pi(i)} \right\Vert^2 + \varepsilon, 
\end{align} 
where in the first inequality the $\varepsilon$ appears by assumption with high probability following \Cref{le.CB^{(M)}easure}, and the second inequality is the triangular inequality. 

We focus now on the terms $\left\vert \hat\alpha_{\pi(i)} - \alpha_{\pi(i)} \right\vert^2$. First of all, we renormalize the coefficients by the factor $(1 - \delta^\prime)$, where $\delta^\prime = m / M$, that is, how many measurements in the initial stage are not contained in the $K$ most relevant bitstrings. It is now easy to bound the error in the estimation of the coefficients as follows. 
\begin{equation}
    \left\vert \hat\alpha_{\pi(i)} - \alpha_{\pi(i)} \right\vert^2 = \left( \Re(\hat\alpha_{\pi(i)} - \alpha_{\pi(i)})\right)^2 + \left( \Im(\hat\alpha_{\pi(i)} - \alpha_{\pi(i)})\right)^2.
\end{equation}
Considering now that we can estimate the values $\Re(\alpha)$ with a Wald estimator~\cite{laplace1820theorie} using $M_\varphi$ using a Hadamard test, then we can bound the error by
\begin{equation}
    \left\vert \hat\alpha_{\pi(i)} - \alpha_{\pi(i)} \right\vert^2 \leq \frac{1}{1 - \delta} \left( \frac{1 - \Re(\hat\alpha_{\pi(i)})^2}{4M_\varphi} + \frac{1 - \Im(\hat\alpha_{\pi(i)})^2}{4M_\varphi}\right) \leq \frac{1}{2M_\varphi}\frac{1}{1- \delta}.
\end{equation}
Since we have $K$ of these terms, then the total error is bounded by
\begin{equation}
    \Vert \ket\psi - \ket{\psi_K}\Vert^2 \leq \varepsilon + \frac{K}{2M_\varphi(1 - \delta)}
\end{equation}

The relationship between the Euclidean distance and the fidelity sets the bounds at
\begin{align}
    \vert \braket{\psi}{\psi_K}\vert^2 \geq & \left(1 - \frac{\varepsilon + \frac{K}{2M_\varphi(1 - \delta)}}{2}\right)^2 = \\
    & 1 - \varepsilon - \frac{K}{2M_\varphi(1 - \delta)} + \frac{\varepsilon^2}{4} + \left( \frac{K}{2M_\varphi(1 - \delta)}\right)^2 + \frac{\varepsilon K}{2M_\varphi(1 - \delta)} \leq \\
    & 1 - \varepsilon - \frac{K}{2M_\varphi(1 - \delta)}
\end{align}

\end{document}